\documentclass[aps,pra,10pt,showpacs,amsmath,longbibliography,twocolumn,floatfix,eqsecnum]{revtex4-1}
\usepackage[dvipsnames]{xcolor}
\usepackage[colorlinks=true,bookmarks=false,linkcolor=NavyBlue,urlcolor=NavyBlue,citecolor=NavyBlue,breaklinks]{hyperref}
\usepackage{amsmath}
\usepackage{amsfonts}
\usepackage{tabularx}
\usepackage{graphicx}
\usepackage{mathtools}
\usepackage{braket}

\newcommand{\real}{\operatorname{Re}}

\newcommand{\parti}[2]{\frac{\partial #1}{\partial #2}}

\newcommand{\Avg}[1]{\left\langle#1\right\rangle}

\newcommand{\abs}[1]{\left|#1\right|}
\newcommand{\bk}[1]{\left(#1\right)}
\newcommand{\Bk}[1]{\left[#1\right]}
\newcommand{\BK}[1]{\left\{#1\right\}}

\newcommand{\expect}{\mathbb E}
\newcommand{\var}{\mathbb V}

\newcommand{\He}{\operatorname{He}}
\newcommand{\floor}[1]{\left\lfloor#1\right\rfloor}
\newcommand{\ceil}[1]{\left\lceil#1\right\rceil}
\newcommand{\cspan}{\operatorname{\overline{span}}}
\begin{document}

\title{Subdiffraction incoherent optical imaging via spatial-mode
  demultiplexing: semiclassical treatment}

\author{Mankei Tsang}
\email{mankei@nus.edu.sg}
\homepage{http://mankei.tsang.googlepages.com/}
\affiliation{Department of Electrical and Computer Engineering,
  National University of Singapore, 4 Engineering Drive 3, Singapore
  117583}

\affiliation{Department of Physics, National University of Singapore,
  2 Science Drive 3, Singapore 117551}

\date{\today}


\begin{abstract}
  I present a semiclassical analysis of a spatial-mode demultiplexing
  (SPADE) measurement scheme for far-field incoherent optical imaging
  under the effects of diffraction and photon shot noise.  Building on
  previous results that assume two point sources or the Gaussian
  point-spread function, I generalize SPADE for a larger class of
  point-spread functions and evaluate its errors in estimating the
  moments of an arbitrary subdiffraction object.  Compared with the
  limits to direct imaging set by the Cram\'er-Rao bounds, the results
  show that SPADE can offer far superior accuracy in estimating the
  second and higher-order moments.
\end{abstract}

\maketitle
\section{Introduction}
Recent theoretical and experimental studies have shown that far-field
optical methods can substantially improve subdiffraction incoherent
imaging
\cite{tnl,sliver,tnl2,nair_tsang16,lupo,tsang16,ant,tsang16c,krovi16,lu16,tham16,tang16,yang16,paur16,rehacek16,yang17,kerviche17,chrostowski17,rehacek17,rehacek17a}. While
most of the prior works focus on two point sources,
Ref.~\cite{tsang16c} proposes a spatial-mode demultiplexing (SPADE)
measurement technique that can enhance the estimation of moments for
arbitrary subdiffraction objects. Although the predicted enhancements
are promising for applications in both astronomy and fluorescence
microscopy, such as size and shape estimation for stellar objects or
fluorophore clusters, researchers in those fields may find it
difficult to comprehend the quantum formalism used in
Ref.~\cite{tsang16c}.  One of the main goals of this work is therefore
to introduce a more accessible semiclassical formalism that can
reproduce the results there, assuming only a background knowledge of
statistical optics on the level of Goodman \cite{goodman,goodman_stat}
and parameter estimation on the level of Van Trees
\cite{vantrees}. The formalism incorporates diffraction, photon shot
noise, and---most importantly---coherent optical processing, which
enables the enhancements proposed in
Refs.~\cite{tnl,sliver,tnl2,nair_tsang16,tsang16,ant,lu16,tsang16c,yang17,rehacek16,kerviche17,tham16,paur16,rehacek17,chrostowski17,lupo,krovi16,tang16,yang16,rehacek17a}.
This treatment thus sheds light on the physical origin of the
enhancements, clarifying that no exotic quantum phenomenon is needed
to explain or implement them.

As Ref.~\cite{tsang16c} assumes the Gaussian point-spread function
(PSF) exclusively, another goal of this work is to generalize the
results for a larger class of PSFs via the theory of orthogonal
polynomials \cite{rehacek16,dunkl}, affirming that enhancements remain
possible in those cases.  To set a benchmark for the proposed method,
I derive limits to moment estimation via direct imaging in the form of
Cram\'er-Rao bounds (CRBs)
\cite{vantrees,zmuidzinas03,huber,feigelson,chao16,diezmann17}, which
are original results in their own right and may be of independent
interest to image-processing research
\cite{brady,villiers,pawley,raginsky10,donoho92,candes14,schiebinger,zhu12,bierbaum17,meister}. On
a more technical level, this work also investigates the estimation
bias introduced by an approximation made in Ref.~\cite{tsang16c} and
assures that it is harmless.

This paper is organized as follows. Section~\ref{sec_formalism}
introduces the background formalism of statistical optics, measurement
noise, and CRBs. Section~\ref{sec_limits} presents the bounds for
moment estimation via direct imaging of a subdiffraction
object. Section~\ref{sec_spade} introduces the theory of SPADE for a
general class of PSFs and evaluates its biases and errors for moment
estimation, showing that giant accuracy enhancements are possible for
the second and higher-order moments. Section~\ref{sec_gauss} revisits
the case of Gaussian PSF studied in Ref.~\cite{tsang16c} and also
proposes new exactly unbiased estimators in the case of two
dimensions. Section~\ref{sec_num} presents a Monte Carlo analysis to
confirm the theory.  Section~\ref{sec_conclusion} concludes the paper,
pointing out open questions and future
directions. Appendices~\ref{sec_mindex}--\ref{sec_finiteM} deal with
mathematical issues that arise in the main text.

\section{\label{sec_formalism}Formalism}
\subsection{Statistical optics}
\begin{figure}[htbp!]
\centerline{\includegraphics[width=\columnwidth]{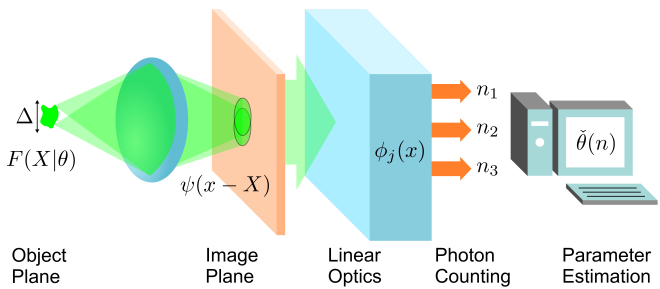}}
\caption{\label{imaging}(Color online). A far-field optical imaging
  system with additional optical processing after the image plane. See
  the main text for the definitions of the expressions.}
\end{figure}

Consider an object emitting spatially incoherent light, a
diffraction-limited imaging system, as depicted in Fig.~\ref{imaging},
and the paraxial theory of quasi-monochromatic scalar waves
\cite{goodman,goodman_stat}.  On the image plane, the mutual coherence
function, also called the mutual intensity, can be expressed as
\cite{goodman,goodman_stat}
\begin{align}
\Gamma(x,x'|\theta)
&= \int dX \psi(x-X)\psi^*(x'-X) F(X|\theta),
\end{align}
where $x, x' \in \mathbb R^D$ are $D$-dimensional position vectors on
the image plane, $X$ is the object-plane position vector normalized
with respect to the magnification factor, $F(X|\theta)$ is the object
intensity function, $\theta = (\theta_0,\theta_1,\dots)$ is a vector
of unknown parameters to be estimated, and $\psi(x)$ is the field PSF.
To simplify the notations, I adopt the multi-index notation described
in Appendix~\ref{sec_mindex} and Ref.~\cite{dunkl}, such that $D$ can
be kept arbitrary, though $D = 1$ or $2$ is typical in spectroscopy
and imaging. Note that three-dimensional imaging requires a different
formalism in the paraxial theory and is outside the scope of this
paper. The mean intensity on the image plane is
\begin{align}
f(x|\theta) &\equiv \Gamma(x,x|\theta) =
\int dX |\psi(x-X)|^2 F(X|\theta),
\label{image}
\end{align}
which is a basic result in statistical optics
\cite{goodman,goodman_stat}.

For convenience, I normalize the position vectors with respect to the
width of the PSF, such that the PSF width is equal to $1$ in this
unit.  The PSF is assumed to obey the normalization
\begin{align}
\int dx |\psi(x)|^2 &= 1,
\label{normF}
\end{align}
such that
\begin{align}
\theta_0 &\equiv \int dX F(X|\theta) = \int dx f(x|\theta)
\end{align}
is the mean optical power reaching the image plane. 

Instead of intensity measurement on the image plane, consider the use
of further linear optics to process the field followed by photon
counting in each output channel, as depicted in
Fig.~\ref{imaging}. The mean power in each output channel can be
expressed as
\begin{align}
p_j(\theta) &= \int dx \int dx'
\phi_j^*(x)\phi_j(x')\Gamma(x,x'|\theta)
\label{p}
\\
&= \int dX \abs{\int dx \phi_j^*(x)\psi(x-X)}^2 F(X|\theta),
\label{p2}
\end{align}
where $\phi_j^*(x)$ is a propagator that couples the image-plane field
from position $x$ to the $j$th output. If the optics after the image
plane is passive, power conservation implies that
\begin{align}
\sum_j p_j(\theta) \le \theta_0.
\label{passive}
\end{align}
This can be satisfied if the set $\{\phi_j(x)\}$ is orthonormal, viz.,
\begin{align}
\int dx \phi_j(x)\phi_k^*(x) &= \delta_{jk},
\label{ortho2}
\end{align}
by virtue of Bessel's inequality \cite{debnath}.  If $\{\phi_j(x)\}$
is also complete in the Hilbert space of image-plane fields, it
becomes an orthonormal basis, and Parseval's identity leads to
equality for Eq.~(\ref{passive}) \cite{debnath}.
Physically, Eq.~(\ref{p}) implies that each output can be regarded as
a projection of the image-plane field in a spatial mode. For example,
direct imaging, which measures the spatial intensity on the image
plane, can be modeled by taking
$\phi_j(x)= \sqrt{dx^{(j)}}\delta(x^{(j)}- x)$, where $x^{(j)}$ is the
position of each pixel with infinitesimal area $dx^{(j)}$, such that
$p_j(\theta) = f(x^{(j)}|\theta) dx^{(j)}$. A generalization of the
measurement model to deal with mode-dependent losses and
non-orthogonal mode projections is possible via the concept of
positive operator-valued measures \cite{tnl2} but not needed here.

In superresolution research, it is known that image processing can
achieve arbitrary resolution if $f(x|\theta)$ is measured exactly and
benign assumptions about the object can be made
\cite{brady,villiers,schiebinger}. The caveat is that the techniques
are severely limited by noise, so the use of proper statistics is
paramount in superresolution studies.  For weak incoherent sources,
such as astronomical optical sources and microscopic fluorophores,
bunching or antibunching is negligible, and it is standard to assume a
Poisson model for the photon counts $n = (n_1,n_2,\dots)$ at the
output channels
\cite{tnl2,goodman_stat,zmuidzinas03,huber,pawley,chao16,vanaert}.
The Poisson distribution is
\begin{align}
P(n|\theta) &= \prod_j \exp\Bk{-\tau p_j(\theta)}
\frac{[\tau p_j(\theta)]^{n_j}}{n_j!},
\label{poisson}
\end{align}
where 
\begin{align}
\tau &\equiv \frac{\eta T}{\hbar\omega},
\end{align} 
$\eta \in [0,1]$ is the detection efficiency, $T$ is the integration
time, and $\hbar\omega$ is the photon energy. The most important
statistics here are the mean
\begin{align}
\expect(n_j) &= \tau p_j(\theta),
\end{align}
where $\expect$ denotes the expectation with respect to $P$, and the
covariance matrix
\begin{align}
\var_{jk}(n) &\equiv \expect(n_jn_k)-\expect(n_j)\expect(n_k)
= \expect\bk{n_j}\delta_{jk},
\label{var}
\end{align}
which is signal-dependent.  If $\{\phi_j\}$ is an orthonormal basis,
the mean photon number detected by the measurement is
\begin{align}
  N &\equiv \sum_j\expect\bk{n_j} =\tau\theta_0.
\label{N}
\end{align}
Conditioned on a total photon number $\sum_j n_j$, $n$ obeys
multinomial statistics, and the reconstruction of $F$ via direct
imaging becomes the density deconvolution problem in nonparametric
statistics; see, for example, Ref.~\cite{meister} and references
therein.

The quantum formalism can arrive at the same Poisson model by assuming
that the source is thermal, the mean photon number per spatiotemporal
mode is much smaller than 1, and the photon count for each channel is
integrated in time over many modes \cite{tnl,tsang16c}.  That said, an
advantage of the semiclassical model besides simplicity is that it
applies to any incoherent source that produces Poisson noise at the
output, such as incoherent laser sources \cite{goodman_stat} and
electron microscopy \cite{vanaert}, without the need to satisfy all
the assumptions of the quantum model.

\subsection{\label{sec_crb}Cram\'er-Rao bounds (CRBs)}
To deal with the signal-dependent nature of Poisson noise, many
existing approaches to computational superresolution
\cite{brady,villiers,donoho92,candes14,schiebinger} are inadequate. A
more suitable tool to derive fundamental limits is the CRB, which is
now standard in astronomy \cite{zmuidzinas03,huber,feigelson} and
fluorescence microscopy \cite{chao16,diezmann17}. For any estimator
$\check\theta(n)$ that satisfies the unbiased condition
\begin{align}
\expect(\check\theta) = \theta,
\end{align}
the mean-square error matrix is equal to its covariance, viz.,
\begin{align}
\textrm{MSE}_{\mu\nu}(\check\theta,\theta) &\equiv 
\expect \bk{\check\theta_{\mu}-\theta_{\mu}}\bk{\check\theta_{\nu}-\theta_{\nu}}
=\var_{\mu\nu}(\check\theta),
\end{align}
and the CRB is 
\cite{vantrees,zmuidzinas03,huber,feigelson,chao16,diezmann17}
\begin{align}
\textrm{MSE}_{\mu\mu}(\check\theta,\theta) &\ge 
\textrm{CRB}_{\mu\mu}(\theta),
\end{align}
where 
\begin{align}
\textrm{CRB}(\theta) &\equiv J^{-1}(\theta)
\end{align}
is the inverse of the Fisher information matrix defined as
\begin{align}
J_{\mu\nu}(\theta) &\equiv \sum_n P(n|\theta)
\parti{\ln P(n|\theta)}{\theta_\mu}
\parti{\ln P(n|\theta)}{\theta_\nu}.
\end{align}
An unbiased estimator whose error attains the CRB is called efficient.
In the limit of infinite trials, the maximum-likelihood estimator is
asymptotically unbiased and efficient \cite{vantrees}, so the bound is
also useful as a measure of the achievable error in the asymptotic
limit.

For the Poisson model, the Fisher information is 
\begin{align}
  J_{\mu\nu}(\theta) &= \tau \sum_j \frac{1}{p_j(\theta)}
\parti{p_j(\theta)}{\theta_\mu}\parti{p_j(\theta)}{\theta_\nu}.
\label{fisher0}
\end{align}
For example, the information for direct imaging with infinitesimal
pixel size is
\begin{align}
J_{\mu\nu}(\theta) &=\tau \int dx \frac{1}{f(x|\theta)}
\parti{f(x|\theta)}{\theta_\mu}
\parti{f(x|\theta)}{\theta_\nu}.
\label{fisher}
\end{align}
The data-processing inequality \cite{zamir} ensures that increasing
the pixel size, or any processing of the image-plane intensity in
general, cannot increase the amount of information.  A simple
extension of Eq.~(\ref{fisher0}) for strong thermal sources with
super-Poisson statistics can be found in Appendix~C of
Ref.~\cite{yang17}.

An intuitive way of understanding Eq.~(\ref{fisher0}) is to regard it
as a signal-to-noise ratio: each derivative
$\partial p_j/\partial\theta_\mu$ measures the sensitivity of an
output to a parameter, while the denominator $p_j$ is proportional to
the Poisson variance and indicates the noise level.  The form of
Eq.~(\ref{fisher0}) hence suggests that any parameter-insensitive
background in $p_j$ should be minimized.  The nonlinear dependence of
the Fisher information on $p_j$ complicates the analysis, but also
hints that coherent optical processing may lead to nontrivial effects.

The Bayesian CRB (BCRB) can be used to set more general limits for any
biased or unbiased estimator
\cite{schutzenberger57,vantrees,gill95,tsang16,bell}. 
Define the Bayesian mean-square error as
\begin{align}
\textrm{BMSE}(\check\theta)
&\equiv \int d\theta \Pi(\theta)\textrm{MSE}(\check\theta,\theta),
\label{BMSE}
\end{align}
where $\Pi(\theta)$ is a prior probability density.  For a prior that
vanishes on the boundary of its domain, the BCRB is 
\begin{align}
\textrm{BMSE}_{\mu\mu}(\check\theta)
&\ge \textrm{BCRB}_{\mu\mu},
\\
\textrm{BCRB} &\equiv \bk{\tilde J + K}^{-1},
\label{BCRB}
\end{align}
where
\begin{align}
\tilde J &\equiv 
\int d\theta \Pi(\theta) J(\theta)
\label{avg_fisher}
\end{align}
is the Fisher information averaged over the prior and
\begin{align}
K_{\mu\nu} &\equiv \int d\theta \frac{1}{\Pi(\theta)} \parti{\Pi(\theta)}{\theta_\mu}
\parti{\Pi(\theta)}{\theta_\nu}
\end{align}
is the prior information.  Other Bayesian bounds for more general
priors can be found in Ref.~\cite{bell}. The BCRB also applies to
the worst-case error $\sup_\theta\textrm{MSE}_{\mu\mu}(\check\theta,\theta)$
for minimax estimation \cite{tsang16,gill95}, since
\begin{align}
\sup_\theta\textrm{MSE}_{\mu\mu}(\check\theta,\theta) &\ge 
\textrm{BMSE}_{\mu\mu}(\check\theta)
\end{align}
for any $\Pi(\theta)$, and the prior can be chosen to tighten the
bound \cite{tsang16,gill95}.

The BCRB is close to the CRB if $J(\theta)$ is constant in the domain
of the prior, such that $\tilde J = J$, and the prior information $K$
is negligible relative to $\tilde J$, such that
\begin{align}
\textrm{BCRB} &= (\tilde J + K)^{-1} \approx \tilde J^{-1} = J^{-1}. 
\end{align}
A counterexample is the problem of two-point resolution
\cite{tsang16}, where $J$ vanishes at a point in the parameter space
and the BCRB becomes very sensitive to the choice of prior, as
mentioned later in Sec.~\ref{sec_special}.

\section{\label{sec_limits}Limits to direct imaging}

\subsection{\label{sec_bounds}Error bounds}
Define the object moments
\begin{align}
\theta_\mu &\equiv \int dX X^\mu F(X|\theta),
&
\mu &\in \mathbb N_0^D,
\label{moment}
\end{align}
as the parameters of interest. Note that the moments are unnormalized,
unlike the definition in Ref.~\cite{tsang16c}. Under general
conditions, the set of moments uniquely determine $F$ \cite{dunkl}, so
there is little loss of generality with this parameterization. I will
focus on moment estimation hereafter and not the pointwise
reconstruction of $F$, however, for two reasons: the moments are more
directly related to many useful parameters in practice, such as the
brightness, location, size, and shape of an object
\cite{nicovich17,feigelson}, while the reconstruction of $F$ without
further prior information is ill-posed and a forlorn task in practice
when noise is present \cite{brady,villiers,donoho92,candes14,meister},
even with the techniques introduced in this work.

Expanding $|\psi(x-X)|^2$ in a Taylor series, the mean image given by
Eq.~(\ref{image}) can be expressed in terms of $\theta$ as
\begin{align}
f(x|\theta) &= \sum_{\mu} \frac{\theta_\mu}{\mu!}(-\partial)^\mu |\psi(x)|^2.
\label{taylor}
\end{align}
The Fisher information given by Eq.~(\ref{fisher}) becomes
\begin{align}
J_{\mu\nu}(\theta) &= \tau 
\int dx \frac{[(-\partial)^\mu|\psi(x)|^2][(-\partial)^\nu|\psi(x)|^2]}
{\mu!\nu! f(x|\theta)}.
\label{Jdirect_exact}
\end{align}
Appendix~\ref{sec_CRBdirect} shows that this can be inverted
analytically to give
\begin{align}
\textrm{CRB}_{\mu\nu}(\theta) &= \frac{\theta_0^2}{N}
\sum_{\xi,\zeta} (C^{-1})_{\mu\xi} M_{\xi\zeta}(\theta) (C^{-1})_{\nu\zeta},
\label{CRBdirect_exact}
\end{align}
where $N$ is the mean photon number given by Eq.~(\ref{N}),
\begin{align}
M_{\mu\nu}(\theta) &\equiv \frac{1}{\theta_0}\int dx f(x|\theta) x^{\mu+\nu}
\label{M}
\end{align}
is the normalized image moment matrix, the $C$ matrix is defined as
\begin{align}
C_{\mu\nu} &\equiv \frac{1}{\nu!}\int dx  |\psi(x)|^2 \partial^\nu x^\mu
\label{C}
\\
&= \left\{\begin{array}{ll}
0, & \textrm{if any}\ \nu_j > \mu_j,
\\
\bk{\begin{array}{c}\mu\\ \nu\end{array}}\Lambda_{\mu-\nu},
& \textrm{otherwise},
\label{C2}
\end{array}
\right.
\end{align}
and
\begin{align}
\Lambda_\mu &\equiv \int dx |\psi(x)|^2 x^\mu
\label{Lambda}
\end{align}
is a moment of the PSF. The lower-triangular property of $C$ indicated
by Eq.~(\ref{C2}) means that $C^{-1}$ is also lower-triangular and the
low-order elements of the CRB can be computed from a finite number of
low-order elements of $M$ and $C$. An unbiased and efficient estimator
is described in Appendix~\ref{sec_est}.

To proceed further, I focus on the subdiffraction regime, which I
define as the scenario where the object support width $\Delta$ is much
smaller than the PSF width. To be specific, the width is defined by
\begin{align}
F(X|\theta) &= 0\ \textrm{if}\ \max_j |X_j| > \frac{\Delta}{2},
\end{align}
and the subdiffraction regime is defined by the condition
\begin{align}
\Delta\ll 1
\end{align}
in the dimensionless unit assumed here. This can be regarded as the
extreme opposite to the sparse regime commonly assumed in compressed
sensing \cite{raginsky10,donoho92,candes14,schiebinger,zhu12} and can
be ensured by prior information in practice. For example, a spot that
resembles the PSF in a prior image indicates a subdiffraction object
and can be studied further via the framework here; such spots are of
course commonly found in both astronomical and microscopic imaging.
In fluorescence microscopy, the subdiffraction support can even be
enforced via stimulated-emission depletion (STED) \cite{hell}, and the
theory here can help STED microscopy gain more information about each
spot beyond $\theta_0$.

In the subdiffraction regime, the moments observe a magnitude
hierarchy with respect to the order $|\mu|$, as
\begin{align}
|\theta_\mu| &\le \int dX |X^\mu| F(X|\theta)
\le \theta_0 \bk{\frac{\Delta}{2}}^{|\mu|},
\label{theta_mag}
\end{align}
and I can combine Eqs.~(\ref{taylor}), (\ref{M}), and (\ref{Lambda})
to obtain
\begin{align}
M_{\mu\nu}(\theta) &= 
\frac{1}{\theta_0}
\sum_{\xi=0}^{\mu+\nu} \theta_\xi \bk{\begin{array}{c}\mu+\nu \\ \xi\end{array}}
\Lambda_{\mu+\nu-\xi}
\label{Mseries}
\\
&= \Lambda_{\mu+\nu} + O(\Delta).
\label{Mpsf}
\end{align}
In other words, the image is so blurred that it resembles the PSF to
the zeroth order, and the image moments approach those of the PSF. The
CRB hence becomes
\begin{align}
\textrm{CRB}_{\mu\nu} &= \frac{\theta_0^2}{N}
\Bk{
\sum_{\xi,\zeta} (C^{-1})_{\mu\xi} \Lambda_{\xi+\zeta} (C^{-1})_{\nu\zeta}
+O(\Delta)}.
\label{CRBdirect}
\end{align}
This is the central result of Sec.~\ref{sec_limits}. 

\begin{figure}[htbp!]
\centerline{\includegraphics[width=\columnwidth]{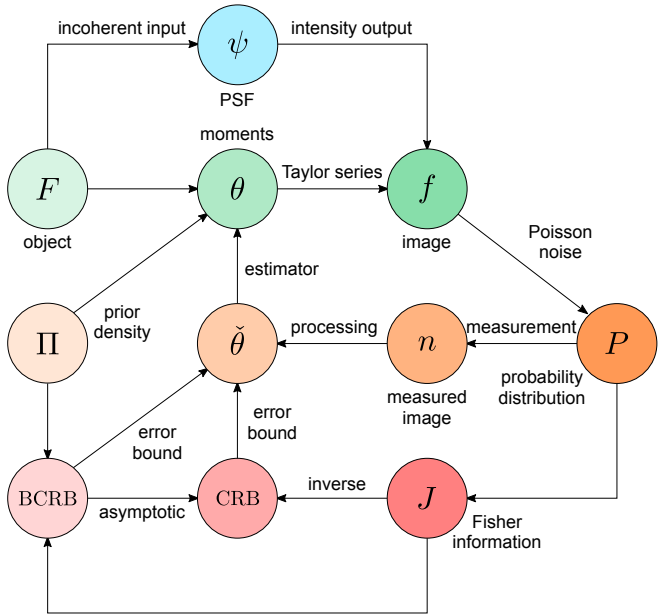}}
\caption{\label{flowchart_direct}(Color online). A flowchart that
  summarizes the relationships among the various quantities defined
  for direct imaging in Sec.~\ref{sec_limits}.}
\end{figure}

To set a more general limit for any biased or unbiased estimator,
consider the BCRB described in Sec.~\ref{sec_crb}.  Since the Fisher
information given by the inverse of Eq.~(\ref{CRBdirect}) depends only
on $\theta_0$ and not the other parameters to the leading order, the
average information $\tilde J$ defined by Eq.~(\ref{avg_fisher}) is
relatively insensitive to the choice of prior in the subdiffraction
regime.  For any reasonable prior that gives a finite prior
information $K$, a long enough integration time can then make
$\tilde J$ much larger than $K$ in Eq.~(\ref{BCRB}), leading to
$\textrm{BCRB} \approx \textrm{CRB}$, if $\theta_0$ is replaced by a
suitable prior value. The two bounds hence give similar results here
in the asymptotic limit.  Figure~\ref{flowchart_direct} summarizes the
relationships among the various quantities defined for direct imaging
in this section.

\subsection{\label{sec_special}Special cases}
The low-order elements of Eqs.~(\ref{Jdirect_exact}) and
(\ref{CRBdirect_exact}) can be used to reproduce a few well known
results. For example, the CRB with respect to $\theta_0$ can be
derived from Eq.~(\ref{CRBdirect_exact}) and is given by
\begin{align}
\textrm{CRB}_{00} = \frac{\theta_0^2}{N},
\end{align}
which is equal to the textbook result.  Another example is
point-source localization \cite{chao16,huber}, for which known results
can be retrieved from Eq.~(\ref{Jdirect_exact}) by defining the
location parameters as $\theta_\mu/\theta_0$ for $|\mu| = |\nu| =
1$. To see this, assume $D = 1$ for simplicity, and the information
with respect to $X = \theta_1/\theta_0$ in the $\Delta \to 0$,
$f(x|\theta) \to \theta_0|\psi(x)|^2$ limit becomes
\begin{align}
J^{(X)} &= \bk{\parti{\theta_1}{X}}^2 J_{11}
\to N  \int dx \frac{[\partial|\psi(x)|^2]^2}{|\psi(x)|^2},
\end{align}
which is exact for one point source \cite{chao16,huber}.

Considering $|\mu| = |\nu| = 2$, Eq.~(\ref{Jdirect_exact}) can also
reproduce the results in Refs.~\cite{tsai79,bettens,vanaert,ram}
regarding sub-Rayleigh two-point separation estimation. To see this,
assume $D = 1$ again and that the centroid of the two point sources is
at the origin. The second moment is then related to the separation $d$
by $\theta_2 = \theta_0 d^2/4$. The information with respect to $d$
becomes
\begin{align}
J^{(d)} &= \bk{\parti{\theta_2}{d}}^2 J_{22} \to
\frac{N d^2}{16} \int dx 
\frac{[\partial^2 |\psi(x)|^2]^2}{|\psi(x)|^2}.
\label{Jd}
\end{align}
This can be compared with a direct calculation of the information by
considering the mean image
\begin{align}
f(x|d) &= \frac{\theta_0}{2} \Bk{|\psi(x-d/2)|^2+|\psi(x+d/2)|^2},
\label{twopoint}
\end{align}
and approximating it for sub-Rayleigh $d\ll 1$ as
\cite{bettens,vanaert,yang17}
\begin{align}
f(x|d) &\approx \theta_0
\Bk{|\psi(x)|^2 + \frac{d^2}{8}\partial^2|\psi(x)|^2}.
\end{align}
The information is then
\begin{align}
J^{(d)} &= \tau \int dx \frac{1}{f}\bk{\parti{f}{d}}^2
\approx \frac{N d^2}{16}
\int dx 
\frac{[\partial^2 |\psi(x)|^2]^2}{|\psi(x)|^2},
\end{align}
which coincides with Eq.~(\ref{Jd}). The vanishing $J^{(d)}$ and
divergent $\textrm{CRB}^{(d)} = 1/J^{(d)}$ for $d \ll 1$ were first
reported in Refs.~\cite{tsai79,bettens,vanaert,ram} and called
Rayleigh's curse in Ref.~\cite{tnl}. The BCRB becomes very sensitive
to the choice of prior and produces a markedly different result from
the CRB when applied to the worst-case error \cite{tsang16}. This
issue depends on the parameterization \cite{gill95} and does not arise
for the moment parameters, however.

In the absence of a specific parametric model or equality parameter
constraints \cite{gorman}, the full information matrix should be
considered, and the CRB given by Eq.~(\ref{CRBdirect_exact}), which
results from inverting the full information matrix, is a tighter limit
\cite{bell} for general objects. Appendix~\ref{sec_nodiff} presents a
limit of Eq.~(\ref{CRBdirect_exact}) when diffraction can be ignored,
while Eq.~(\ref{CRBdirect}) should be used in the subdiffraction
regime.

This section has established fundamental limits to direct imaging in
the subdiffraction and shot-noise-limited regime. The next sections
show that coherent optical processing can beat them.

\section{\label{sec_spade}Spatial-mode demultiplexing (SPADE)}

\subsection{\label{sec_pad}Point-spread-function-adapted (PAD) basis}
References~\cite{tnl,sliver,tnl2,nair_tsang16,tsang16,ant,lu16,tsang16c,yang17,rehacek16,kerviche17,tham16,paur16,rehacek17,chrostowski17,lupo,krovi16,tang16,yang16,rehacek17a}
have shown that SPADE, a technique of linear optics and photon
counting with respect to a judiciously chosen basis of spatial modes,
can substantially improve subdiffraction imaging. To generalize the
use of the TEM basis in Ref.~\cite{tsang16c}, I consider the
point-spread-function-adapted (PAD) basis proposed by Rehacek
\textit{et al.}\ for the two-point problem \cite{rehacek16} and apply
it to more general objects. Denote the PAD basis by
\begin{align}
\BK{\phi_q(x); q \in \mathbb N_0^D},
\label{PAD_set}
\end{align}
where the spatial modes are more conveniently defined in the
spatial-frequency domain. Defining
\begin{align}
\Phi_q(k) &\equiv \frac{1}{(2\pi)^{d/2}}\int dk \phi_q(x) \exp(-ik\cdot x),
\label{Phi}
\\
\Psi(k)  &\equiv \frac{1}{(2\pi)^{d/2}}\int dk \psi(x) \exp(-ik\cdot x),
\label{Psi}
\end{align}
$\Phi_q(k)$ can be expressed as
\begin{align}
\Phi_q(k) &= (-i)^{|q|}g_q(k) \Psi(k),
\label{Phi_pad}
\\
g_q(k) &\equiv \sum_r G_{qr} k^r,
\label{cpoly}
\end{align}
where $\{g_q(k); q \in \mathbb N_0^D\}$ is a set of real orthogonal
polynomials with $|\Psi(k)|^2$ as the weight function
\cite{dunkl}, $G$ is an invertible matrix that satisfies the
lower-triangular property
\begin{align}
  G_{qr} &= 0\ \textrm{if}\ r > q,
\label{lower_tri_C}
\end{align}
and the indices follow a total and degree-respecting order that obeys
\begin{align}
r \ge q \Rightarrow |r| \ge |q|.
\end{align}
See Appendix~\ref{sec_CRBdirect} for more details about orthogonal
polynomials. The polynomials are assumed to satisfy the orthonormal
condition
\begin{align}
\int dk \Phi_q^*(k)\Phi_r(k) &= \int dk |\Psi(k)|^2 g_q(k) g_r(k) = \delta_{qr},
\label{cortho}
\end{align}
which also ensures that $\{\phi_q\}$ is orthonormal.  The completeness of
$\{\phi_q\}$ can be proved along the lines of Ref.~\cite{dunkl} but is
not essential here. As $\phi_0(x) = \psi(x)$ and each higher-order
mode in real space is a sum of $\psi(x)$ derivatives given by
\begin{align}
\phi_q(x) = (-i)^{|q|}g_q(-i\partial)\psi(x),
\end{align}
the PAD basis can be regarded as a generalization of the binary SPADE
concept in Ref.~\cite{tnl} and the derivative-mode concept in
Ref.~\cite{paur16}.

In terms of the PAD basis, I can define a mutual coherence matrix as
\begin{align}
\Gamma_{qq'}(\theta) &\equiv \int dX h_q(X)h_{q'}^*(X) F(X|\theta),
\label{Gamma}
\\
h_q(X) &\equiv \int dx \phi_q^*(x)\psi(x-X).
\label{h}
\end{align}
In particular, SPADE in terms of the PAD basis gives a set of output
channels with powers
\begin{align}
p_q(\theta) &= \int dX |h_q(X)|^2 F(X|\theta) = 
\Gamma_{qq}(\theta),
\label{p_pad}
\end{align}
and the Poisson photon counts $\{n_q; q \in \mathbb N_0^D\}$ have
expected values
\begin{align}
  \expect\bk{n_q} &= \tau_0 p_q(\theta),
\label{mbar_pad}
&
\tau_0 &\equiv \frac{\eta_0 T}{\hbar\omega},
\end{align}
where $\eta_0$ is the efficiency of the PAD-basis measurement.
An unbiased estimator of $\Gamma_{qq}$ is
\begin{align}
\check \Gamma_{qq} &= \frac{n_q}{\tau_0},
\label{est_pad}
\end{align}
and its variance is
\begin{align}
\var\bk{\check \Gamma_{qq}} &= \frac{\Gamma_{qq}}{\tau_0}.
\label{var_pad}
\end{align}

In the context of the Gaussian PSF, Refs.~\cite{yang16,tsang16c} found
that $p_q(\theta)$ is sensitive only to some of the object moments. To
estimate the other moments, Ref.~\cite{tsang16c} further proposes
measurements that access the off-diagonal elements of $\Gamma$.  To
measure an off-diagonal $\Gamma_{qq'}$, take two spatial modes with
indices $q$ and $q'$ from the PAD basis and interfere them, such that
the outputs correspond to projections into the spatial modes
\begin{align}
\varphi_{qq'}^+(x) &= 
\frac{1}{\sqrt{2}}\Bk{\phi_q(x)+\phi_{q'}(x)},
\\
\varphi_{qq'}^-(x) &= \frac{1}{\sqrt{2}}\Bk{\phi_q(x)-\phi_{q'}(x)},
\end{align}
which I call interferometric-PAD (iPAD) modes.  The 
powers at the two outputs are
\begin{align}
p_{qq'}^+ &= 
\frac{\Gamma_{qq} +  \Gamma_{q'q'}}{2} + \real\Gamma_{qq'},
\label{pplus}
\\
p_{qq'}^- &=
\frac{\Gamma_{qq} +  \Gamma_{q'q'}}{2} - \real\Gamma_{qq'}.
\label{pminus}
\end{align}
The photon counts, denoted by $n_{qq'}^+$ and
$n_{qq'}^-$, have expected values
\begin{align}
\expect\bk{n_{qq'}^+} &= \tau_s p_{qq'}^+,
&
\expect\bk{n_{qq'}^-} &= \tau_s p_{qq'}^-,
\label{mbar_ipad}
&
\tau_s &\equiv \frac{\eta_s T}{\hbar\omega},
\end{align}
where $\eta_s$ denotes the efficiency of the measurement that includes
these two projections.  
Assume further that $|\Psi(k)|^2$ is centrosymmetric, as defined by
\begin{align}
|\Psi(k)|^2 &= |\Psi(-k)|^2,
\label{even}
\end{align}
such that $G$, $h_q(X)$, and $\Gamma_{qq'}$ are all real, as shown in
Appendix~\ref{sec_matrices2} and assumed hereafter. An unbiased
estimator of $\Gamma_{qq'}$ is then
\begin{align}
\check \Gamma_{qq'} &= \frac{n_{qq'}^+- n_{qq'}^-}{2\tau_s},
\label{est_ipad}
\end{align}
with
\begin{align}
\var\bk{\check \Gamma_{qq'}} &= \frac{\Gamma_{qq}+ \Gamma_{q'q'}}{4\tau_s}.
\label{var_ipad}
\end{align}
The estimators $\check\Gamma_{qq'}$ given by Eqs.~(\ref{est_pad}) and
(\ref{est_ipad}) will be used in Sec.~\ref{sec_moment} to construct
moment estimators.

Since the iPAD modes are not orthogonal to the PAD modes, they cannot
belong to the same orthonormal basis.  This means that, if projections
into both PAD and iPAD modes are desired, multiple measurements in
different bases are needed and must be performed on different photons.
This can be done either sequentially in time via configurable
interferometers or on different beamsplitted parts of the light. If
each measurement has an efficiency $\eta_s$, energy conservation
mandates that
\begin{align}
\sum_s \eta_s \le 1.
\label{conserve}
\end{align}

\subsection{\label{sec_moment}Moment estimation}

To relate $\Gamma$ to the object moments, use
Eqs.~(\ref{Phi})--(\ref{Phi_pad}) to rewrite the propagator $h_q(X)$
in Eq.~(\ref{h}) as
\begin{align}
h_q(X) &= i^{|q|}\int dk |\Psi(k)|^2g_q(k) \exp(-ik\cdot X)
\label{h_k}
\\
&= i^{|q|}
\int dk |\Psi(k)|^2g_q(k)\sum_{r}\frac{(-ik)^r X^r}{r!}
\\
&= \sum_{r} H_{qr} X^r,
\label{hpoly}
\end{align}
where
\begin{align}
H_{qr} &\equiv  \frac{i^{|q|}}{r!}\int dk |\Psi(k)|^2 g_q(k) (-ik)^r
\label{H}
\\
&= \frac{i^{|q|}(-i)^{|r|}}{r!}(G^{-1})_{rq},
\label{HC}
\\
(H^{-1})_{qr} &= q! i^{|q|}(-i)^{|r|} G_{rq},
\label{Hinv}
\end{align}
as shown in Appendix~\ref{sec_matrices2}. Since $G^{-1}$ and $G$ are
lower-triangular, $H$ and $H^{-1}$ are upper-triangular, satisfying
\begin{align}
H_{qr} = 0,\ (H^{-1})_{qr} = 0\ \textrm{if}\ r < q.
\label{weak_H}
\end{align}
Substituting Eq.~(\ref{hpoly})
into Eq.~(\ref{Gamma}), $\Gamma_{qq'}$ can be related to the moments by
\begin{align}
\Gamma_{qq'} &= \sum_{r,r'} H_{qr}H_{q'r'}\theta_{r+r'},
\label{Gamma_theta}
\end{align}
which shows that each $\Gamma_{qq'}$ is sensitive to a combination of
moments with orders at least as high as $|q+q'|$.  Given the
magnitudes of $\theta$ according to Eq.~(\ref{theta_mag}), the
magnitude of $\Gamma_{qq'}$ can be expressed as
\begin{align}
\Gamma_{qq'} &= \theta_0 O(\Delta^{|q+q'|}),
\label{Gmag}
\end{align}
and the variances of the estimators given by Eqs.~(\ref{var_pad}) and
(\ref{var_ipad}) become
\begin{align}
\var\bk{\check \Gamma_{qq'}} &= \frac{\theta_0^2}{N_s} O(\Delta^{2\min(|q|,|q'|)}),
\label{var_mag}
  \\
  N_s &\equiv \tau_s\theta_0 = \frac{\eta_s T\theta_0}{\hbar\omega}.
\end{align}
Equations~(\ref{Gmag}) and (\ref{var_mag}) will be used to evaluate
the errors of moment estimation.

Instead of computing the CRB and relying on asymptotic
arguments, here I construct explicit moment estimators and evaluate
their errors directly to demonstrate the achievable performance of
SPADE. To begin, consider the inverse of Eq.~(\ref{Gamma_theta}) given
by
\begin{align}
\theta_{q+q'} &= \sum_{r,r'} 
(H^{-1})_{qr}(H^{-1})_{q'r'}\Gamma_{rr'},
\label{theta_Gamma}
\end{align}
which implies that an unbiased estimator of $\theta_{q+q'}$ can be
constructed from unbiased estimators of $\Gamma_{pp'}$ given by
Eqs.~(\ref{est_pad}) and (\ref{est_ipad}), viz.,
\begin{align}
\check\theta_{q+q'} &= \sum_{r,r'} 
(H^{-1})_{qr}(H^{-1})_{q'r'}
\check\Gamma_{rr'}.
\label{theta_est}
\end{align}
This estimator may not be realizable, however, as it may not be
possible to group the needed projections into a reasonable number of
bases.  A fortuitous exception occurs for the Gaussian PSF, as
elaborated later in Sec.~\ref{sec_unbiased}.

\begin{figure}[htbp!]
\centerline{\includegraphics[width=\columnwidth]{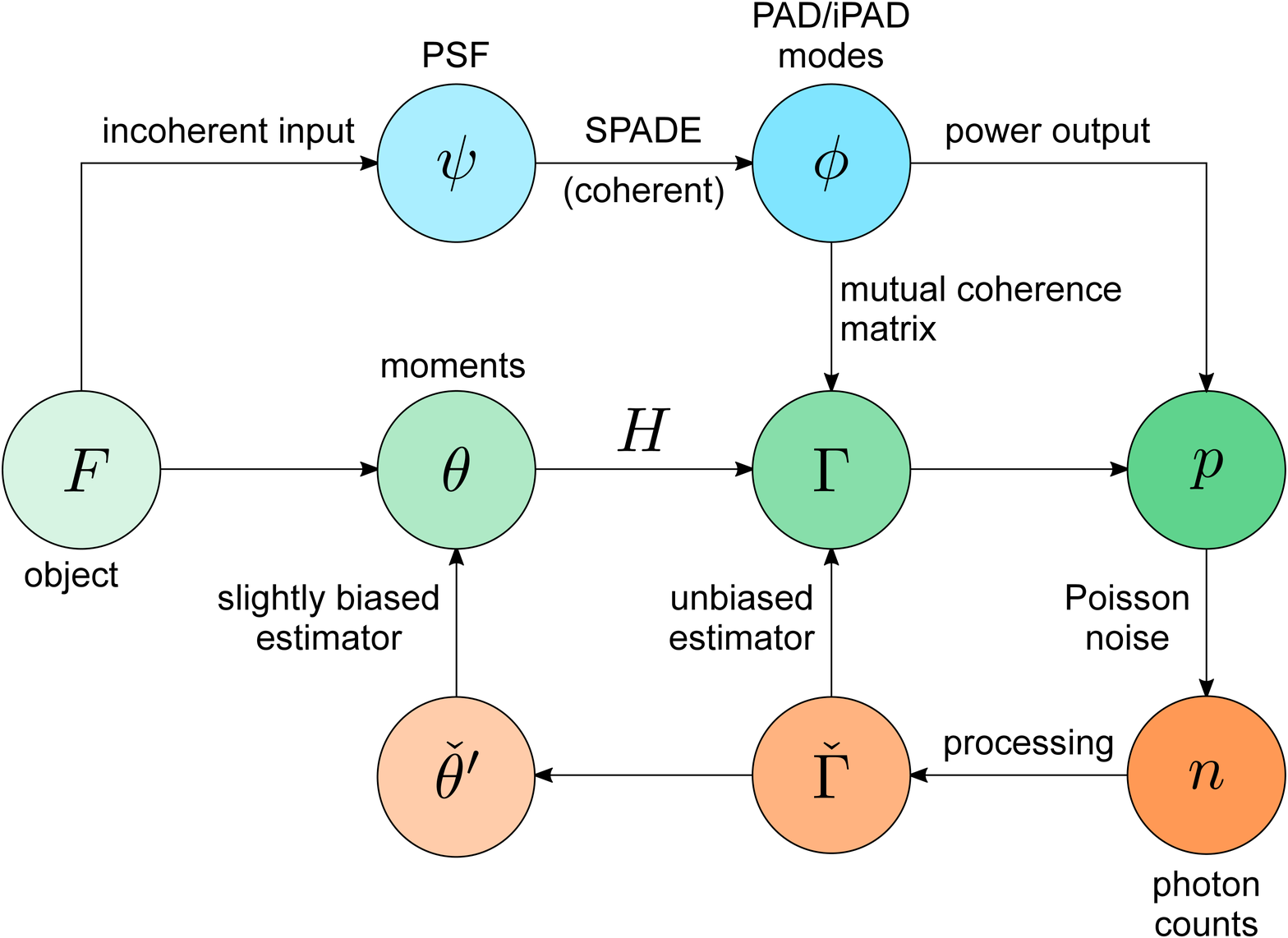}}
\caption{\label{flowchart_spade}(Color online). A flowchart that
  summarizes the relationships among the various quantities defined
  for SPADE in Sec.~\ref{sec_spade}.}
\end{figure}

To find a simpler estimator, I focus on the class of separable PSFs
given by
\begin{align}
|\Psi(k)|^2 &= \prod_j |\Psi^{(j)}(k_j)|^2,
\end{align}
where each $|\Psi^{(j)}(k_j)|^2$ is a one-dimensional function.  
Defining
\begin{align}
g_{q_j}^{(j)}(k_j) &= \sum_{r_j} G_{q_j r_j}^{(j)} k_j^{r_j}
\end{align}
as the orthogonal polynomials with respect to each
$|\Psi^{(j)}(k_j)|^2$, the natural orthogonal polynomials in the
multivariate case are their products, viz.,
\begin{align}
g_q(k) &= \prod_j g_{q_j}^{(j)}(k_j).
\end{align}
As each $G_{q_jr_j}^{(j)}$ is lower-triangular, I obtain the condition
\begin{align}
G_{qr} &= \prod_jG_{q_j r_j}^{(j)} = 0\ 
\textrm{if any}\ r_j > q_j.
\end{align}
It follows from Eqs.~(\ref{HC}) and (\ref{Hinv}) that $H$ and $H^{-1}$
are also separable and given by
\begin{align}
H_{qr} &= \prod_j\frac{i^{q_j}(-i)^{r_j}}{r_j!}\Bk{(G^{(j)})^{-1}}_{r_j q_j},
\\
\bk{H^{-1}}_{qr} &= \prod_jq_j!i^{q_j}(-i)^{r_j} G^{(j)}_{r_j q_j}.
\end{align}
Using the property
\begin{align}
(H^{-1})_{qr} &= 0\ \textrm{if any}\ q_j > r_j,
\label{strong_H}
\end{align}
I can rewrite the sums in Eq.~(\ref{theta_Gamma}) as
\begin{align}
\sum_{r} &= \sum_{r_1 = q_1}^\infty  \dots
\sum_{r_D=q_D}^\infty
\end{align}
and obtain
\begin{align}
\theta_{q+q'} &= (H^{-1})_{qq}(H^{-1})_{q'q'}\Gamma_{qq'}
\nonumber\\&\quad
+ 
\sum_{|r+r'| > |q+q'|} (H^{-1})_{qr}(H^{-1})_{q'r'}  \Gamma_{rr'},
\end{align}
which consists of one $\theta_0 O(\Delta^{|q+q'|})$ term and
higher-order terms, as ranked by Eq.~(\ref{Gmag}).  To evaluate the
magnitude of the higher-order terms, note that, for a centrosymmetric
$|\Psi(k)|^2$, $(H^{-1})_{qr} \propto G_{rq} = 0$ if
$|r| = |q| + 1, |q|+3,\dots$ \cite{dunkl}, so
\begin{align}
\sum_{|r+r'|>|q+q'|} (H^{-1})_{qr}(H^{-1})_{q'r'}  \Gamma_{rr'}
&=\theta_0 O(\Delta^{|q+q'|+2}),
\label{theta_high}
\end{align}
which is smaller than the leading-order term by two orders of
magnitude. A simplified estimator, involving only one
$\check\Gamma_{qq'}$, can then be constructed as
\begin{align}
\check\theta_{q+q'}' &= (H^{-1})_{qq}(H^{-1})_{q'q'}\check \Gamma_{qq'}
= \frac{\check\Gamma_{qq'}}{H_{qq}H_{q'q'}},
\label{sim_est}
\end{align}
where the last step uses the fact $(H^{-1})_{qq} = 1/H_{qq}$ for a
triangular matrix. The bias is then the negative of
Eq.~(\ref{theta_high}), viz.,
\begin{align}
\expect\bk{\check\theta_{q+q'}'}-\theta_{q+q'} &= 
\theta_0 O(\Delta^{|q+q'|+2}).
\label{bias_mag}
\end{align}
Figure~\ref{flowchart_spade} summarizes the relationships
among the various quantities defined in this section, while
Appendix~\ref{sec_nonsep} discusses a generalization of the estimator
for non-separable PSFs.

Given Eq.~(\ref{var_mag}), the variance of the estimator is
\begin{align}
\var\bk{\check\theta_{q+q'}'} &= 
\frac{\var\bk{\check \Gamma_{qq'}}}{H_{qq}^2H_{q'q'}^2}
= \frac{\theta_0^2}{N_s}O(\Delta^{2\min(|q|,|q'|)}).
\label{var_moment}
\end{align}
To minimize the variance for a given moment $\theta_\mu$ with
$\mu = q+q'$, $\min(|q|,|q'|)$ should be made as high as
possible. This can be accomplished by choosing
\begin{align}
&\textrm{for each}\ j \in \BK{1,2,\dots,D},
\nonumber\\
&q_j = \left\{
\begin{array}{ll}
\mu_j/2 & \textrm{if}\ \mu_j\ \textrm{is even},
\\
\floor{\mu_j/2} & \textrm{if}\ \mu_j\ \textrm{is the first odd number},
\\
\ceil{\mu_j/2} & \textrm{if}\ \mu_j\ \textrm{is odd and the last
choice was}\ \floor{},
\\
\floor{\mu_j/2} & \textrm{if}\ \mu_j\ \textrm{is odd and the last
choice was}\ \ceil{}.
\end{array}
\right.
\label{afloor}
\end{align}
The alternating floor ($\floor{}$) and ceil ($\ceil{}$) operations
keep $|q|$ high without exceeding $|q'|$.  If $|\mu|$ is even, $\mu$
has an even number of odd elements, then $|q| = |q'| = |\mu|/2$. If
$|\mu|$ is odd, $\mu$ has an odd number of odd elements, then
$|q| = (|\mu|-1)/2$ and $|q'| = (|\mu|+1)/2$.  Hence one can achieve
\begin{align}
\min(|q|,|q'|) &= \floor{\frac{|\mu|}{2}},
\\
\var\bk{\check\theta_{\mu}'} &= \frac{\theta_0^2}{N_s}
O(\Delta^{2\floor{|\mu|/2}}),
\label{var_theta}
\end{align}
and the mean-square error becomes
\begin{align}
\textrm{MSE}(\check\theta_\mu',\theta_\mu) &= \var\bk{\check\theta_\mu'}
+\Bk{\expect\bk{\check\theta_\mu'}-\theta_\mu}^2
\\
&= \frac{\theta_0^2}{N_s}
O(\Delta^{2\floor{|\mu|/2}}) + \theta_0^2 O(\Delta^{2|\mu|+4}).
\label{error_moment}
\end{align}
Compared with the CRB for direct imaging given by
Eq.~(\ref{CRBdirect}), Eq.~(\ref{error_moment}) can be much lower in
the $\Delta \ll 1$ subdiffraction regime if $|\mu| \ge 2$, the bias is
negligible, and $\eta_s$ is on the same order of magnitude as the
direct-imaging efficiency.  This is the central result of
Sec.~\ref{sec_spade}. The conclusion holds also from the Bayesian or
minimax perspective, since the BCRB for direct imaging is close to the
CRB in the asymptotic limit, as argued in Sec.~\ref{sec_bounds}, while
Eq.~(\ref{error_moment}) also applies to the Bayesian or worst-case
error for SPADE if $\theta_0$ is replaced by a suitable prior value.

A heuristic explanation of the enhancements is as follows. Recall that
Poisson noise is signal-dependent, and any background in the signal
increases the variance.  In the subdiffraction regime, the direct
image is so blurred that it resembles the PSF $|\psi(x)|^2$, and the
fundamental mode $\phi_0(x) = \psi(x)$ acts as a background and the
main contributor of noise.  With SPADE, on the other hand, each moment
estimator is designed to use spatial modes with the highest possible
orders. The isolation from the lower-order modes, including the
fundamental, substantially reduces the background and improves the
signal-to-noise ratio.

\subsection{\label{sec_multi}Multi-moment estimation}

\begin{figure*}[htbp!]
  \centerline{\includegraphics[width=\textwidth]{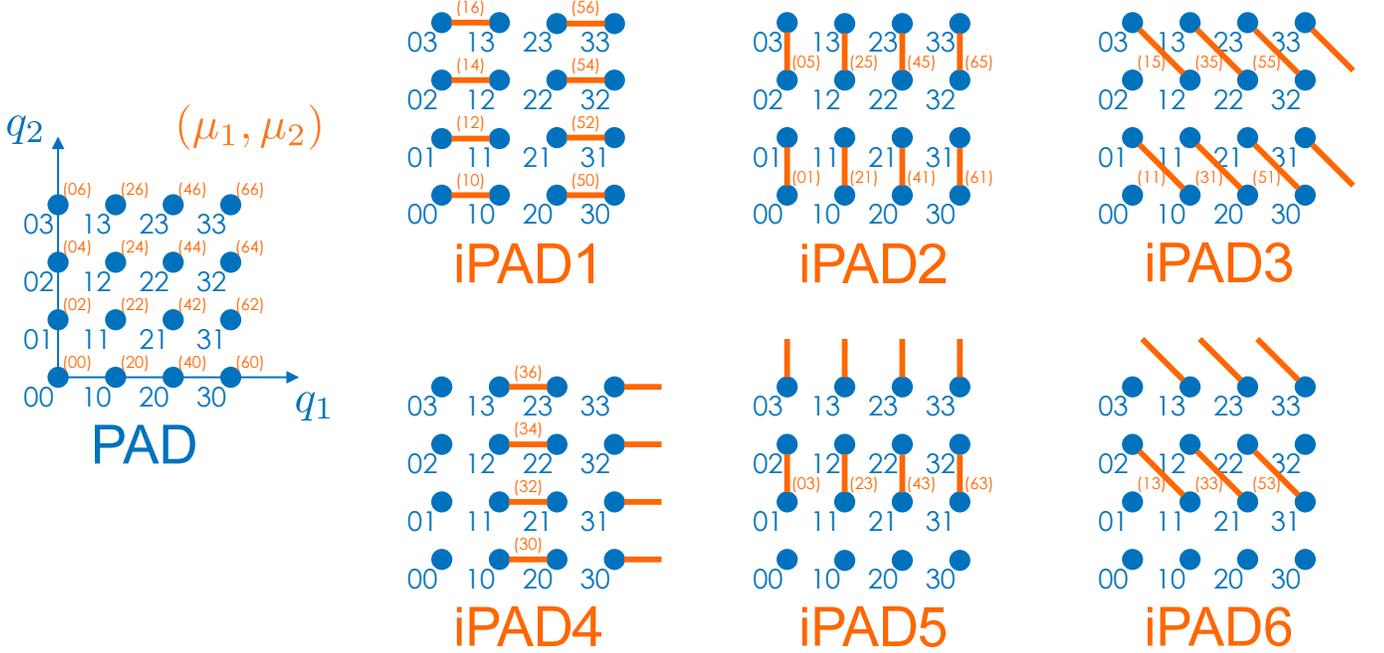}}
  \caption{\label{ipads}(Color online). An illustration of the PAD and
    iPAD1--iPAD6 bases in the mode-index space. Each dot in the
    $(q_1,q_2)$ space represents a PAD mode, and the PAD modes form
    the PAD basis on the left.  For each iPAD basis, a line connecting
    two dots represents an interference between the two PAD modes,
    producing two new modes that replace the original PAD modes in the
    basis. Each bracketed pair of numbers denote the order
    $(\mu_1,\mu_2) = (q_1+q_1',q_2+q_2')$ of the estimator
    $\check\theta_\mu'$ that a projection can provide via
    Eq.~(\ref{sim_est}). In each iPAD basis, the unconnected dots
    represent the PAD modes that complete the basis and can also be
    measured to provide extra information.}
\end{figure*}

The remaining question is the number of bases needed to estimate all
moments. For $D = 1$, three bases are enough: a measurement in the PAD
basis provides
\begin{align}
\BK{\check\Gamma_{qq}, q \in \mathbb N_0}\ \textrm{and}\ 
\BK{\check\theta_\mu'; \mu \in 2\mathbb N_0},
\end{align}
where $2\mathbb N_0=\{0,2,4,\dots\}$,
a measurement in the basis
$\{\varphi_{q,q+1}^\pm(x); q \in 2\mathbb N_0\}$ provides
\begin{align}
\BK{\check\Gamma_{q,q+1}, q \in 2\mathbb N_0}\ \textrm{and}\ 
\BK{\check\theta_\mu'; \mu \in 4\mathbb N_0 + 1},
\end{align}
where $4\mathbb N_0 + 1 = \{1,5,9,\dots\}$, and a measurement in the basis
$\{\varphi_{q,q+1}^\pm(x); q \in 2\mathbb N_0+1\}$ provides
\begin{align}
\BK{\check\Gamma_{q,q+1}, q \in 2\mathbb N_0+1}\ \textrm{and}\ 
\BK{\check\theta_\mu'; \mu \in 4\mathbb N_0 + 3},
\end{align}
where $2\mathbb N_0+1 = \{1,3,5,\dots\}$ and
$4\mathbb N_0 + 3 = \{3,7,11,\dots\}$. If the light is split for
measurements in all three basis, the condition of energy conservation
given by Eq.~(\ref{conserve}) implies
\begin{align}
\min(\eta_s) &\le \frac{1}{3}.
\end{align}
For $D = 2$, seven bases---defined by Table~\ref{projections} and
illustrated by Fig.~\ref{ipads}---can do the job. I call these bases
PAD and iPAD1--iPAD6, which generalize the TEM and iTEM1--iTEM6 bases
proposed in Ref.~\cite{tsang16c} for the Gaussian PSF.  Energy
conservation now implies
\begin{align}
  \min(\eta_s) &\le \frac{1}{7},
\end{align}
if measurements in all the seven bases are performed. The essential
point is that the penalty in efficiency for multi-moment estimation is
only a constant factor, and significant enhancements over direct
imaging remain possible.

\begin{table*}[htbp!]
\centerline{
\begin{tabularx}{0.7\textwidth}{|l|l|X|X|X|X|}
  \hline Basis & Spatial modes & $q_1$ & $q_2$
  & $\mu_1 = q_1+q_1'$ & $\mu_2 = q_2+q_2'$\\
  \hline
  PAD & $\phi_q(x)$ & $\mathbb N_0$ & $\mathbb N_0$ & $2\mathbb N_0$ & $2\mathbb N_0$ \\
  iPAD1 & $\varphi_{qq'}^\pm(x); q' = q+(1,0)$
  & $2\mathbb N_0$ & $\mathbb N_0$ & $4\mathbb N_0 + 1$ & $2\mathbb N_0$ \\
  iPAD2 & $\varphi_{qq'}^\pm(x); q' = q+(0,1)$
  & $\mathbb N_0$ & $2\mathbb N_0$ &  $2\mathbb N_0$ & $4\mathbb N_0 + 1$\\
  iPAD3 & $\varphi_{qq'}^\pm(x); q' = q+(1,-1)$
  & $\mathbb N_0$ & $2\mathbb N_0 + 1$  &  $2\mathbb N_0 + 1$ & $4\mathbb N_0 + 1$\\
  iPAD4 & $\varphi_{qq'}^\pm(x); q' = q+(1,0)$
  & $2\mathbb N_0 + 1$ & $\mathbb N_0$ & $4\mathbb N_0 + 3$ & $2\mathbb N_0$\\
  iPAD5 & $\varphi_{qq'}^\pm(x); q' = q+(0,1)$
  & $\mathbb N_0$ & $2\mathbb N_0 + 1$ & $2\mathbb N_0$ & $4\mathbb N_0 + 3$\\
  iPAD6 & $\varphi_{qq'}^\pm(x); q' = q+(1,-1)$
  & $\mathbb N_0$ & $2\mathbb N_0+2$ & $2\mathbb N_0 + 1$ & $4\mathbb N_0 + 3$\\
  \hline
\end{tabularx}
}
\caption{A list of measurement bases for moment estimation with
  a $D = 2$ separable PSF and their spatial modes. Measurement in
  each basis can provide a set of moment estimators $\{\check\theta_\mu'\}$
  according to Eq.~(\ref{sim_est}),
  where the set of $\mu = (\mu_1,\mu_2)$ indices are listed in the last two columns.
  The case of $D = 1$ can be retrieved by considering
  the PAD, iPAD1, and iPAD4 bases and $q_1$ and $\mu_1$ only.}
\label{projections}
\end{table*}

\subsection{\label{sec_inform}Criterion for informative estimation}
A word of caution is in order: even with SPADE, there are severe
resolution limits.  This is because the moments are inherently small
parameters in the subdiffraction regime according to
Eq.~(\ref{theta_mag}), and the error needs be much smaller than the
prior range of the parameter for the estimation to be informative.  To
evaluate the usefulness of an estimation relative to prior
information, I adopt the Bayesian perspective
\cite{vantrees,bell,berger} and consider the Bayesian error given by
Eq.~(\ref{BMSE}). In the absence of measurements, the error is
determined by the prior and given by
\begin{align}
\textrm{BMSE}_{\mu\mu}^{(\Pi)} &\equiv 
\expect^{(\Pi)}\Bk{\theta_\mu-\expect^{(\Pi)}(\theta_\mu)}^2 
\le \theta_0^2\bk{\frac{\Delta}{2}}^{2|\mu|},
  \label{prior_error}
\end{align}
where $\expect^{(\Pi)}$ denotes the expectation with respect to
$\Pi(\theta)$, the upper bound comes from Eq.~(\ref{theta_mag}), and
$\theta_0$ is assumed to be given for simplicity.  Using the bound as
a conservative estimate of the prior error, a rule of thumb for
informative estimation is
\begin{align}
\frac{\textrm{BMSE}_{\mu\mu}}{\theta_0^2(\Delta/2)^{2|\mu|}} \ll 1.
\end{align}
The small prior error places a stringent requirement on the
post-measurement error.  For direct imaging, assuming the asymptotic
limit where the BCRB is close to the CRB given by
Eq.~(\ref{CRBdirect}), the fractional BCRB is
\begin{align}
\frac{\textrm{BCRB}_{\mu\mu}}{\textrm{BMSE}_{\mu\mu}^{(\Pi)}} &\approx
\frac{\textrm{CRB}_{\mu\mu}}{\textrm{BMSE}_{\mu\mu}^{(\Pi)}} 
=  \frac{O(\Delta^{-2|\mu|})}{N}.
\end{align}
This value grows exponentially with the order $|\mu|$, meaning that
the estimation of higher-order moments requires exponentially more
photons to become informative.

For SPADE, an achievable Bayesian error can be obtained by averaging
$\textrm{MSE}(\check\theta_\mu',\theta_\mu)$, and the magnitude is
also given by Eq.~(\ref{error_moment}). The fractional error becomes
\begin{align}
\frac{\textrm{BMSE}_{\mu\mu}}
{\textrm{BMSE}_{\mu\mu}^{(\Pi)}}  &= 
\frac{O(\Delta^{2\floor{|\mu|/2}-2|\mu|})}{N_s} +  O(\Delta^{4}).
\end{align}
The $O(\Delta^4)$ relative bias is always much smaller than $1$, but
the fractional variance still grows with $|\mu|$
exponentially. Compared with direct imaging, the exponent is reduced
for $|\mu| \ge 2$ and not as many photons are needed to achieve a
small fractional error for a given moment, but higher-order moments
remain more difficult to estimate.

This consideration suggests that SPADE is most useful for scenarios
that depend on only a few low-order moments.  For example, the
two-point problem studied in
Refs.~\cite{tnl,sliver,tnl2,nair_tsang16,tsang16,ant,lu16,yang17,kerviche17,tham16,paur16,rehacek16,chrostowski17,lupo,krovi16,tang16,yang16,rehacek17a}
requires moments up to the second order only \cite{tsang16c}, the case
of two unequal sources studied in Refs.~\cite{rehacek17,rehacek17a}
requires moments up to the third, and parametric object models with
size and shape parameters \cite{tsang16c,bierbaum17} can also be
related to low-order moments.

\section{\label{sec_gauss}Gaussian point-spread function}
\subsection{Direct imaging}
For an illustrative example of the general theory, consider the
Gaussian PSF
\begin{align}
\psi(x) &= \frac{1}{(2\pi)^{d/4}}\exp\bk{-\frac{||x||^2}{4}},
\end{align}
which is a common assumption in fluorescence microscopy
\cite{chao16,zhang07}.  The Hermite polynomials can be used to compute
the CRB in the limit of $\Delta \to 0$, as shown in
Appendix~\ref{sec_CRBgauss}.  The result is
\begin{align}
\textrm{CRB}_{\mu\nu} &\to \frac{\theta_0^2}{N}\mu!\delta_{\mu\nu},
\label{CRBgauss}
\end{align}
which coincides with the $D = 2$ theory in Ref.~\cite{tsang16c}.

\subsection{SPADE}
The PSF in the spatial-frequency domain is
\begin{align}
\Psi(k) &= \bk{\frac{2}{\pi}}^{d/4}\exp(-||k||^2).
\end{align}
A set of orthogonal polynomials with respect to $|\Psi(k)|^2$ are
defined by
\begin{align}
g_q(k) &= \frac{1}{\sqrt{q!}} \He_q(2k),
\end{align}
and the PAD mode functions become
\begin{align}
\Phi_q(k)
&= \bk{\frac{2}{\pi}}^{d/4}\frac{(-i)^{|q|}}{\sqrt{q!}} \He_q(2k)\exp(-||k||^2),
\\
\phi_q(x) &= \frac{1}{(2\pi)^{d/4}\sqrt{q!}}\He_q(x) \exp\bk{-\frac{||x||^2}{4}}.
\end{align}
The PAD basis in this case is simply the TEM basis, as expected.  The
propagator given by Eq.~(\ref{h}) can be computed analytically with
the help of the generating function for Hermite polynomials
\cite{NIST:DLMF,Olver:2010:NHMF}; the result is
\begin{align}
h_q(X) &= H_{qq}\exp\bk{-\frac{||X||^2}{8}} X^q,
\label{h_gauss}
\\
H_{qq} &= \frac{1}{2^{|q|}\sqrt{q!}}.
\end{align}
The mutual coherence matrix $\Gamma$ defined by Eq.~(\ref{Gamma})
becomes
\begin{align}
 \Gamma_{qq'} &= H_{qq}H_{q'q'}\int dX
\exp\bk{-\frac{||X||^2}{4}} X^{q+q'}F(X|\theta).
\label{Gamma_gauss}
\end{align}
Unbiased estimators of $\Gamma_{qq'}$ can be constructed from
projections in the PAD and iPAD spatial modes according to
Eqs.~(\ref{est_pad}) and (\ref{est_ipad}); the iPAD modes are called
iTEM modes in Ref.~\cite{tsang16c}. The estimator variances are given
by Eqs.~(\ref{var_pad}) and (\ref{var_ipad}), with magnitudes given by
Eq.~(\ref{var_mag}).

To estimate a given moment $\theta_\mu$, $q$ and $q' = \mu-q$ can be
chosen according to Eq.~(\ref{afloor}), the simplified estimator given
by Eq.~(\ref{sim_est}) can be used, and the error then agrees with
Eq.~(\ref{error_moment}).  These results again agree with
Ref.~\cite{tsang16c}, except that Ref.~\cite{tsang16c} neglects the
contribution of bias to the mean-square error and therefore does not
include the second term in Eq.~(\ref{error_moment}).

\subsection{\label{sec_unbiased}Exactly unbiased estimator}
For $D = 2$, the PAD and iPAD1--iPAD6 bases described by
Table~\ref{projections} and Fig.~\ref{ipads} become the TEM and
iTEM1--iTEM6 bases proposed in Ref.~\cite{tsang16c}, and the estimator
given by Eq.~(\ref{sim_est}) is equivalent to the ones
proposed in Ref.~\cite{tsang16c}.  Interestingly, it is possible to go
further than Ref.~\cite{tsang16c} and construct exactly unbiased
moment estimators from these measurements. First note that
Eq.~(\ref{Gamma_gauss}) offers a shortcut to express each moment in
terms of $\Gamma$ as follows:
\begin{align}
\theta_{q+q'} &= \int dX \exp\bk{\frac{||X||^2}{4}}
\exp\bk{-\frac{||X||^2}{4}} X^{q+q'} F(X|\theta)
\\
&= 
\int dX \sum_r \frac{X^{2r}}{r! 4^{|r|}}
\exp\bk{-\frac{||X||^2}{4}} X^{q+q'} F(X|\theta)
\\
&= \sum_r \frac{1}{r! 4^{|r|}}
\int dX 
\exp\bk{-\frac{||X||^2}{4}} X^{q+q'+2r} F(X|\theta)
\\
&= \sum_r \frac{ \Gamma_{q+r,q'+r}}{r! 4^{|r|} H_{q+r,q+r}H_{q'+r,q'+r}}.
\label{theta_gauss}
\end{align}
Combining Eqs.~(\ref{sim_est}) and (\ref{theta_gauss}), it can then be
shown that the estimator
\begin{align}
\check\theta_\mu &= \sum_r \frac{\check\theta_{\mu+2r}'}{r!4^{|r|}}
\end{align}
is exactly unbiased. To construct 
\begin{align}
\BK{\check\theta_\mu; \mu \in (2\mathbb N_0) \times (2\mathbb N_0)},
\end{align}
one simply needs
$\BK{\check\theta_\mu'; \mu \in (2\mathbb N_0) \times (2\mathbb
  N_0)}$ from the PAD basis. To construct
\begin{align}
\BK{\check\theta_\mu; \mu \in (2\mathbb N_0+1) \times (2\mathbb N_0)},
\end{align}
one needs
$\BK{\check\theta_\mu'; \mu \in (2\mathbb N_0+1) \times (2\mathbb
  N_0)}$, which can be obtained from the iPAD1 and iPAD4 bases.
Similarly, to construct
\begin{align}
\BK{\check\theta_\mu; \mu \in (2\mathbb N_0)\times (2\mathbb N_0+1)},
\end{align}
one needs
$\BK{\check\theta_\mu'; \mu \in (2\mathbb N_0)\times (2\mathbb
  N_0+1)}$, which can be obtained from the iPAD2 and iPAD5 bases.
Finally, to construct
\begin{align}
\BK{\check\theta_\mu; \mu \in (2\mathbb N_0+1)\times (2\mathbb N_0+1)},
\end{align}
one needs
$\BK{\check\theta_\mu'; \mu \in (2\mathbb N_0+1)\times (2\mathbb
  N_0+1)}$, which can be obtained from the iPAD3 and iPAD6 bases.  The
error matrix of the unbiased estimator becomes
\begin{align}
\textrm{MSE}_{\mu\nu}(\check\theta,\theta) &= \var_{\mu\nu}(\check\theta)
  =\frac{\theta_0^2}{\min(N_s)}O(\Delta^{2\floor{|\mu|/2}}) \delta_{\mu\nu},
\end{align}
which remains on the same order of magnitude as the variance of the
simplified estimator in Eq.~(\ref{error_moment}), while the bias
contribution is no longer present. The number of bases needed to
achieve enhanced and exactly unbiased multi-moment estimation for
other PSFs and dimensions remains an open question.

\section{\label{sec_num}Numerical demonstration}
\begin{figure}[htbp!]
\centerline{\includegraphics[width=\columnwidth]{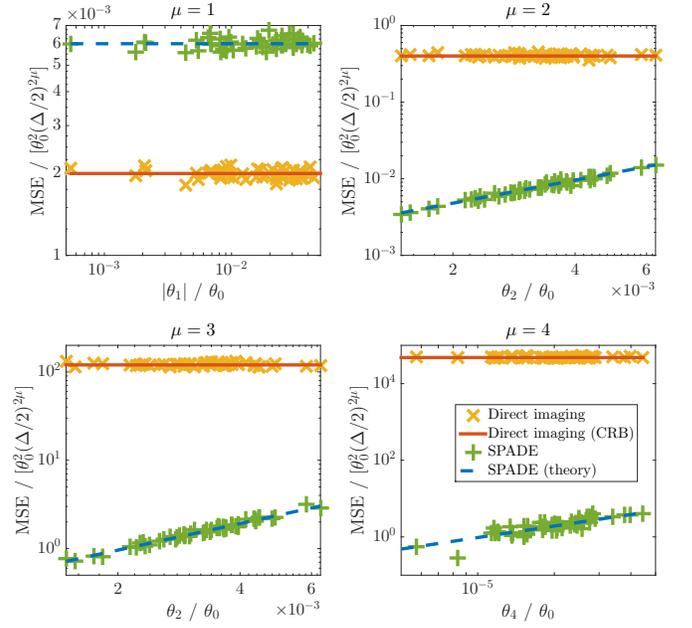}}
\caption{\label{gauss_sim}(Color online). Simulated and theoretical
  errors of moment estimation via direct imaging and SPADE for the
  Gaussian PSF. The discrete points are from the Monte Carlo
  simulations and the lines are from the analytic theory.  All axes
  are normalized, dimensionless, and in logarithmic scale. See the main
  text for details.}
\end{figure}

\begin{figure}[htbp!]
\centerline{\includegraphics[width=\columnwidth]{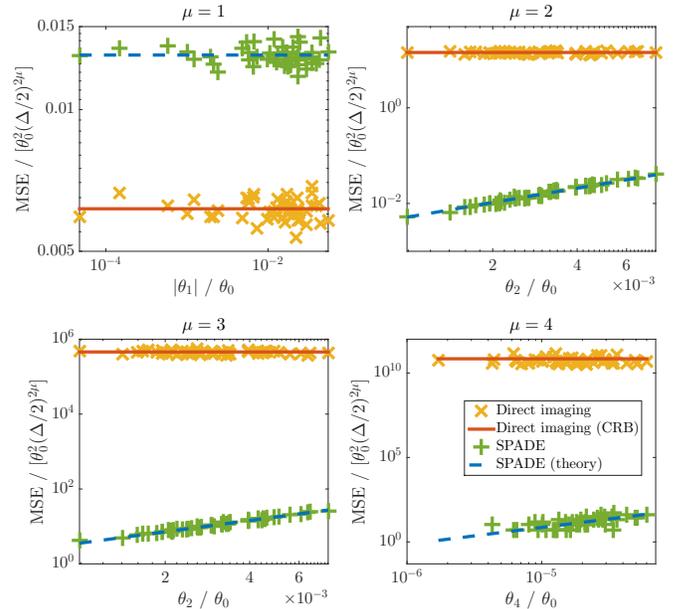}}
\caption{\label{bump_sim}(Color online). Simulated and theoretical
  errors of moment estimation via direct imaging and SPADE for the
  bump aperture given by Eq.~(\ref{bump}). The format of the plots is
  identical to that of Fig.~\ref{gauss_sim}. See the main text for
  details.  }
\end{figure}

I now present Monte Carlo simulations to corroborate the theory.
Assume $D = 1$. Each simulated object is an ensemble of $S = 5$ point
sources with randomly generated positions
$\{X_\sigma; \sigma = 1,\dots,S\}$ within the interval
\begin{align}
|X_\sigma| &\le \frac{\Delta}{2}, & \Delta &= 0.2,
\end{align}
such that
\begin{align}
F(X|\theta) &= \frac{\theta_0}{S}\sum_{\sigma=1}^S \delta(X-X_\sigma).
\end{align}
$50$ objects are generated for each PSF under study.  For direct
imaging, I assume that the mean photon number is $N = 50,000$, the
pixel size is $dx = 0.1$, and $1,000$ samples of Poisson images are
generated for each object. The estimator described in
Appendix~\ref{sec_est} is applied to each sample to estimate the
moments $\theta_\mu$ for $\mu = 1,2,3,4$ ($\theta_0$ can be estimated
by summing all the photon counts and the results are trivial). The
sample errors with respect to the true parameters are averaged to
approximate the expected values.  The averaged errors are then plotted
for two different PSFs in Figs.~\ref{gauss_sim} and \ref{bump_sim} and
compared with the CRB given by Eq.~(\ref{CRBdirect}), omitting the
$O(\Delta)$ correction. 

To simulate SPADE according to Sec.~\ref{sec_spade}, measurements in
three different bases are simulated. The first basis is
\begin{align}
\BK{\phi_0(x),\phi_1(x),\phi_2(x)},
\end{align}
with the simulated photon counts denoted by $\{n_0,n_1,n_2\}$, the
second basis is
\begin{align}
\BK{\varphi_{01}^+(x),\varphi_{01}^-(x),\phi_2(x)},
\end{align}
with the photon counts denoted by $\{n_{01}^+,n_{01}^-,n_2'\}$,
and the third basis is
\begin{align}
\BK{\phi_0(x),\varphi_{12}^+(x),\varphi_{12}^-(x)},
\end{align}
with the photon counts denoted by $\{n_{0}',n_{12}^+,n_{12}^-\}$. The
light is split equally among the three measurements, such that
$N_s = N/3$. All photons in higher-order modes are neglected.

\begin{figure}[htbp!]
\centerline{\includegraphics[width=\columnwidth]{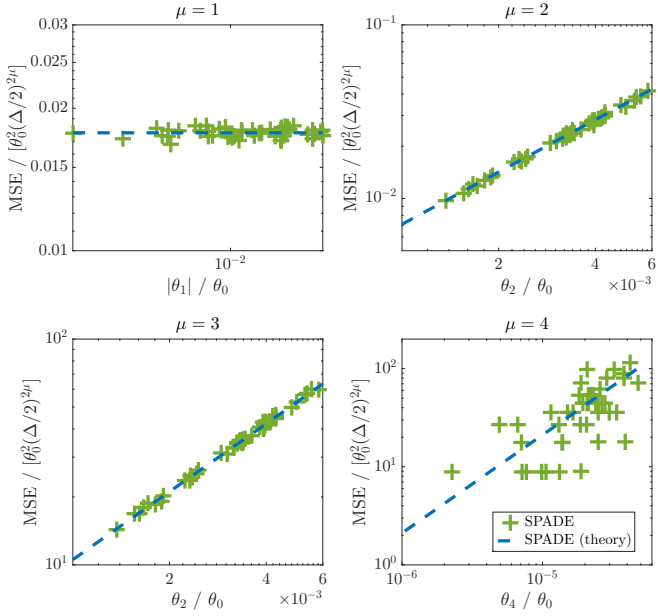}}
\caption{\label{hard_sim}(Color online). Simulated and theoretical
  errors of moment estimation via SPADE for the rectangle aperture
  given by Eq.~(\ref{hard}). The format of the plots is identical to
  that of Figs.~\ref{gauss_sim} and \ref{bump_sim}.  Note that the
  spread of errors for $\mu = 4$ looks more severe because the range
  of the vertical axis is smaller than those of the other $\mu = 4$
  plots in Figs.~\ref{gauss_sim} and \ref{bump_sim}. Also, relatively
  few photons from a subdiffraction object are coupled into the
  $\phi_2(x)$ mode, so the error itself has a high variance and more
  samples would be required for the average errors to get closer to
  the expected values.}
\end{figure}

\begin{figure}[htbp!]
\centerline{\includegraphics[width=\columnwidth]{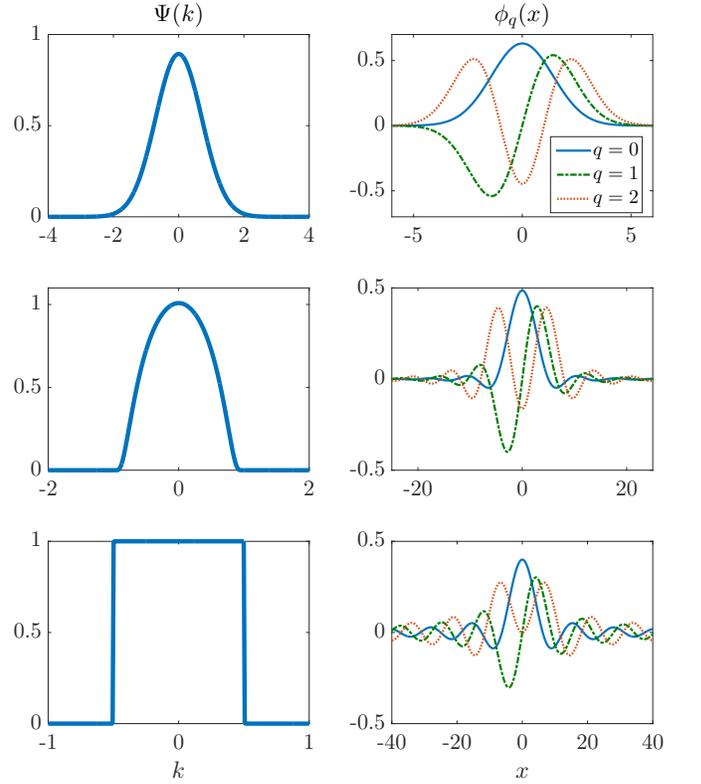}}
\caption{\label{psf_pad}(Color online). Left column: the aperture
  functions for the three PSFs under study: the Gaussian (first row),
  the bump given by Eq.~(\ref{bump}) (second row), and the rectangle
  given by Eq.~(\ref{hard}) (third row).  Right column: the PAD modes
  $\phi_0(x)$, $\phi_1(x)$, and $\phi_2(x)$ for each PSF.  All the
  axes follow the dimensionless units defined in the main text.}
\end{figure}

To estimate the moments with SPADE, I use the simplified but biased
estimator given by Eq.~(\ref{sim_est}), with $q$ given by
Eq.~(\ref{afloor}).  Using Eq.~(\ref{est_ipad}) for
$\check\Gamma_{01}$, the estimator of $\theta_1$ becomes
\begin{align}
\check\theta_{1}' &= \frac{\check\Gamma_{01}}{H_{00}H_{11}}
= \frac{n_{01}^+-n_{01}^-}{2H_{00}H_{11}\tau_s}.
\end{align}
The estimator is applied to $1,000$ samples of the simulated photon
counts for each object.  The sample errors with respect to the true
parameters are averaged and compared with the analytic expression
\begin{align}
\textrm{MSE}_{11} &\approx 
\var\bk{\check\theta_1'} 
=\frac{\var\bk{\check\Gamma_{01}}}{H_{00}^2H_{11}^2}
\approx  \frac{\Gamma_{00}}{4H_{00}^2H_{11}^2\tau_s}
\approx \frac{\theta_0}{4H_{11}^2\tau_s},
\end{align}
which neglects the bias and applies
the approximations
\begin{align}
\Gamma_{qq} + \Gamma_{q'q'} \approx \Gamma_{qq} \approx H_{qq}^2\theta_{2q}
\end{align}
to Eqs.~(\ref{var_ipad}) and (\ref{Gamma_theta}). Similarly,
\begin{align}
\check\theta_2' &= \frac{n_1}{H_{11}^2\tau_s},
&
\textrm{MSE}_{22} &\approx  \frac{\theta_2}{H_{11}^2 \tau_s},
\\
\check\theta_3' &= \frac{n_{12}^+-n_{12}^-}{2H_{11}H_{22}\tau_s},
&
\textrm{MSE}_{33} &\approx \frac{\theta_2}{4H_{22}^2\tau_s}.
\end{align}
To estimate $\theta_4$, I use both of the photon counts that come from
the two $\phi_2(x)$ projections to obtain
\begin{align}
\check\theta_4' &= \frac{n_2+n_2'}{2H_{22}^2\tau_s},
&
\textrm{MSE}_{44} &\approx \frac{\theta_4}{2H_{22}^2\tau_s}.
\end{align}
There is no need to specify $\theta_0$, $\tau$, or $\tau_s$
individually if the errors are normalized with respect to
$\theta_0^2$. The simulated errors and the analytic expressions are plotted in
Figs.~\ref{gauss_sim}--\ref{hard_sim} against the relevant parameters
in log-log scale for the three PSFs.
The three PSFs in the spatial-frequency domain under study
and the associated PAD modes are plotted in Fig.~\ref{psf_pad}.  

Figure~\ref{gauss_sim} plots the results for the Gaussian PSF
described in Sec.~\ref{sec_gauss}. The simulated errors all match the theory,
despite the approximations in the analytic expressions. In particular,
the agreement confirms that the contribution of bias to the errors of
SPADE is negligible. For $\mu = 1$, SPADE uses one third of the
photons only, and its errors are three times those of direct
imaging. For higher moments, however, SPADE outperforms direct imaging
by orders of magnitude.

It is important to note that the plotted mean-square errors are
normalized with respect to $\theta_0^2(\Delta/2)^{2\mu}$, which is the
square of the prior limit given by Eq.~(\ref{theta_mag}), and only the
normalized errors for $\mu =1,2$ go significantly below $1$. According
to the discussion in Sec.~\ref{sec_inform}, this implies that only the
estimation for $\mu \le 2$ is informative, while the estimation for
$\mu \ge 3$ would require a lot more photons to become informative.
The high variances of the estimators for $\mu \ge 3$ also suggest
that, for the given photon number, replacing them with Bayesian
estimators \cite{vantrees,bell,berger} can reduce their errors to the
vicinity of the prior levels given by Eq.~(\ref{prior_error}),
although the bias will go up a lot.

The second PSF under study is the ``bump'' aperture function
\cite{debnath}
\begin{align}
\Psi(k) &= \left\{
\begin{array}{ll}
\Psi(0) \exp\bk{-\frac{k^2}{1-k^2}}, & |k| < 1,
\\
0, & |k| \ge 1,
\end{array}
\right.
\label{bump}
\end{align}
where $\Psi(0) \approx 1.0084$ is a normalization constant.  The
compact support models a hard bandwidth limit, while the infinite
differentiability of $\Psi(k)$ ensures that all the moments of
$|\psi(x)|^2$ are finite and the direct-imaging theory in
Sec.~\ref{sec_limits} is valid, as discussed in
Appendix~\ref{sec_finiteM}. The simulated errors, plotted in
Fig.~\ref{bump_sim}, behave similarly to those in the Gaussian case,
except that the direct-imaging errors are substantially higher for
higher moments.  The enhancements by SPADE appear even bigger, though
not big enough to bring the errors for $\mu \ge 3$ down to the
informative regime for the given photon number.

The final PSF is the textbook rectangle aperture function
\begin{align}
\Psi(k) &= \left\{
\begin{array}{ll}
1, & |k| < 1/2,
\\
0, & |k| \ge 1/2.
\end{array}
\right.
\label{hard}
\end{align}
The second and higher moments of $|\psi(x)|^2$ are infinite, meaning
that the direct-imaging theory in Sec.~\ref{sec_limits} is
inapplicable, as discussed in Appendix~\ref{sec_finiteM}. Fortunately,
the orthogonal polynomials with respect to $|\Psi(k)|^2$ and therefore
the PAD basis remain well-defined
\cite{rehacek16}. Figure~\ref{hard_sim} plots the results for SPADE,
which are similar to those for the bump aperture in
Fig.~\ref{bump_sim}.  Although these results have no direct-imaging
limits to compare with, the earlier results on the two-point problem
for this PSF \cite{tnl,paur16,rehacek16} suggest that significant
improvements remain likely.

\section{\label{sec_conclusion}Conclusion}
The semiclassical treatment complements the quantum approach in
Ref.~\cite{tsang16c} by offering a shortcut to the Poisson
photon-counting model for incoherent sources, passive linear optics,
and photon counting. Besides pedagogy, this work generalizes the
results in
Refs.~\cite{tnl,sliver,tnl2,nair_tsang16,lupo,tsang16,ant,tsang16c,krovi16,lu16,tham16,tang16,yang16,paur16,rehacek16,yang17,kerviche17,chrostowski17,rehacek17,rehacek17a}
for more general objects and PSFs in the context of moment estimation,
demonstrating that the giant enhancements by SPADE are not limited to
the case of two point sources or Gaussian PSF considered in prior
works.

Many open problems remain, such as extensions for more general PSFs,
more complex objects, and three-dimensional imaging, the effect of
excess statistical and systematic errors, such as dark counts,
aberrations, turbulence, and nonparaxial effects \cite{mortensen10},
the application of more advanced Bayesian or minimax statistics
\cite{vantrees,bell,raginsky10,donoho92,candes14,schiebinger,zhu12,bierbaum17,meister},
and the quantum optimality of the measurements
\cite{tnl,nair_tsang16,lupo,tsang16,ant,tsang16c,krovi16,lu16,rehacek16,chrostowski17,rehacek17,rehacek17a}. Experimental
implementation is another important future direction.  For
proof-of-concept demonstrations, it should be possible to use the same
setups described in Refs.~\cite{tang16,yang16,tham16,paur16} to
estimate at least the second moments of more general objects.  For
practical applications in astronomy and fluorescence microscopy,
efficient demultiplexing for broadband sources is needed.  The
technical challenge is by no means trivial, but the experimental
progress on spatial-mode demultiplexers has been encouraging
\cite{tham16,tang16,paur16,yang16,morizur,li14,luo14,mohanty17,martin17,mirhosseini13,zhou17},
and the promise of giant imaging enhancements using simply far-field
linear optics should motivate further efforts.

\section*{Acknowledgments}
This work is supported by the Singapore Ministry of Education Academic
Research Fund Tier 1 Project R-263-000-C06-112.

\appendix
\section{\label{sec_mindex}Multi-index notation}
A $D$-dimensional vector of continuous variables is written as
\begin{align}
 x &= (x_1,x_2,\dots,x_D) \in \mathbb R^D.
\end{align}
For such a vector, the following notations are assumed:
\begin{align}
dx &\equiv \prod_{j=1}^D dx_j,
&
\int dx &\equiv \int_{\mathbb R^D} dx,
\nonumber\\
\delta(x-x') &\equiv \prod_{j=1}^D \delta(x_j-x_j'),
&
\partial_x &\equiv \bk{\parti{}{x_1},\dots,\parti{}{x_D}},
\nonumber\\
k\cdot x &\equiv \sum_{j=1}^D k_jx_j,
&
||x||^2 &\equiv x\cdot x.
\end{align}
If the subscript is omitted in $\partial$, derivatives with respect to
$x$ are assumed.

A vector of integer indices, on the other hand, is defined as
\begin{align}
\mu &= (\mu_1,\mu_2,\dots,\mu_D) \in \mathbb N_0^D.
\end{align}
For such a vector, the following notations are assumed:
\begin{align}
0 &\equiv \bk{0,\dots,0},
&
|\mu| &\equiv \sum_{j=1}^D |\mu_j|,
\nonumber\\
\sum_\mu &\equiv \sum_{\mu \in \mathbb N_0^D},
&
\sum_{\mu=\nu}^\xi &\equiv \sum_{\mu=\nu_1}^{\xi_1}
\dots \sum_{\mu=\nu_D}^{\xi_D},
\nonumber\\
\mu! &\equiv \prod_{j=1}^D \mu_j!.
&
\bk{\begin{array}{c}\mu\\ \nu\end{array}}
&\equiv \frac{\mu!}{(\mu-\nu)!\nu!}.
\end{align}
Note that the one-norm is assumed for index vectors.  Other useful
notations include
\begin{align}
x^\mu &\equiv \prod_{j=1}^D x_j^{\mu_j},
&
\partial_x^\mu &\equiv \prod_{j=1}^D \parti{^{\mu_j}}{x_j^{\mu_j}}.
\end{align}

\section{\label{sec_CRBdirect}CRB for direct imaging}
It is useful to define a Hilbert space
\begin{align}
\mathcal H &\equiv \cspan\BK{b_\mu(x); \mu \in \mathbb N_0^D}
\end{align}
with respect to
\begin{align}
b_\mu(x) &\equiv \frac{(-\partial)^\mu |\psi(x)|^2}{\mu!\tilde f(x|\theta)},
&
\tilde f(x|\theta) &\equiv \frac{f(x|\theta)}{\theta_0},
\end{align}
and the weighted inner product
\begin{align}
\Avg{u,v} &\equiv \int dx \tilde f(x|\theta) u(x)v(x),
\end{align}
where $\cspan$ is the closed linear span inside the $L^2(\tilde f)$
space \cite{dunkl,debnath} and $\tilde f(x|\theta)$ is the normalized
image.  In other words, any function in $\mathcal H$ can be expressed
as a linear combination of $\{b_\mu(x)\}$.
Equation~(\ref{Jdirect_exact}) becomes
\begin{align}
J_{\mu\nu} &= \frac{\tau}{\theta_0}\Avg{b_\mu,b_\nu}.
\label{Jdirectb}
\end{align}
This can be inverted with the help of orthogonal polynomials.  Define
\begin{align}
a &\equiv \BK{a_{\mu}(x); \mu \in \mathbb N_0^D},
\end{align}
where $a_\mu(x)$ is a real polynomial with degree $|\mu|$ and the
orthonormal condition is
\begin{align}
\Avg{a_\mu,a_\nu} = \delta_{\mu\nu}.
\label{ortho}
\end{align}
For orthogonal polynomials to exist, the moment matrix $M$ given by
Eq.~(\ref{M}) should be positive-definite \cite{dunkl}, or
equivalently
\begin{align}
\int dx \tilde f(x|\theta) \mathcal P^2(x) > 0
\end{align} 
for any polynomial $\mathcal P$. The strict positiveness can be
satisfied as long as the support of $\tilde f(x|\theta)$ is an
infinite set, as $\mathcal P^2(x)$ has a finite number of zeros only.

The orthogonal polynomials can be computed by applying the
Gram-Schmidt procedure to the set of monomials
$\{x^\mu; \mu \in \mathbb N_0^D\}$ if the set is totally ordered
\cite{dunkl}.  For $D = 1$, the natural order $\{1,x,x^2,\dots\}$
leads to a unique set of orthogonal polynomials for a given weight
function. For $D \ge 2$, however, the situation is more complicated.
A useful requirement is that the order should respect the degree in
the sense of
\begin{align}
\nu \ge \mu \Rightarrow |\nu| \ge |\mu|.
\label{respect}
\end{align}
An example is the graded lexicographical order, defined by
\begin{align}
\nu  > \mu
&\Leftrightarrow
|\nu| > |\mu|, \textrm{or if } |\nu| = |\mu|,
\nonumber\\
&\quad
\textrm{ the first nonzero }\nu_j-\mu_j > 0.
\label{glex}
\end{align}
For $D = 2$ for example, the order is
\begin{align}
(0,0) &< \nonumber\\
 (0, 1) &< (1,0) < 
\nonumber\\
(0,2) &< (1,1) < (2,0) < \dots
\nonumber\\
(0, |\mu|) &< (1,|\mu|-1) < \dots < (|\mu|, 0) < \dots,
\end{align}
but one should see in this example that indices with the same total
degree $|\mu|$ may be ordered in other ways and there is no single
compelling choice; a different choice will lead to a different set of
orthogonal polynomials. In the following I assume simply that a
degree-respecting order has been chosen; the analysis is valid
regardless of the choice.

Express each polynomial as
\begin{align}
a_\mu(x) &= \sum_{\nu} A_{\mu\nu} x^\nu,
\label{apoly}
\end{align}
where $A$ is a matrix that satisfies the lower-triangular property
\begin{align}
A_{\mu\nu} = 0\ \textrm{if}\ \nu > \mu.
\end{align}
Combining Eqs.~(\ref{M}), (\ref{ortho}), and (\ref{apoly}),
I obtain
\begin{align}
\sum_{\xi,\zeta} A_{\mu\xi}M_{\xi\zeta}A_{\nu\zeta} &= \delta_{\mu\nu}.
\end{align}
Given a total order of the indices, the matrices can be rasterized
into two-dimensional matrices. Equation~(\ref{Aortho}) can then be
written more compactly as
\begin{align}
A M A^\top &= I,
\label{Aortho}
\end{align}
where $\top$ denotes the matrix transpose and $I$ is the identity
matrix. As $M$ is positive-definite, $A$ can be obtained from the
Cholesky decomposition
\begin{align}
M = L L^\top,
\end{align}
where $L$ is a real lower-triangular matrix with positive diagonal
elements \cite{horn}.  Since the diagonal elements of a triangular
matrix are also its eigenvalues, $L$ is invertible, $L^{-1}$ is also
lower-triangular, and setting
\begin{align}
A = L^{-1}
\end{align}
leads to
\begin{align}
M &= (A^{-1})(A^{-1})^\top,
\label{MA}
\end{align}
which satisfies Eq.~(\ref{Aortho}).

To invert Eq.~(\ref{Jdirectb}), I also need to prove that $a$ is an
orthonormal basis in $\mathcal H$.  The orthonormality given by
Eq.~(\ref{ortho}) is satisfied by definition, while the completeness
follows from the fact that the only function
$u(x) = \sum_\nu \lambda_\nu b_\nu(x)$ in $\mathcal H$ that is
orthogonal to $a$ in the sense of
\begin{align}
\Avg{a_\mu,u} &= \sum_\nu \Avg{a_\mu,b_\nu}\lambda_\nu  = 0,
&
\mu &\in \mathbb N_0^D,
\end{align}
is the zero function, provided that 
\begin{align}
B_{\mu\nu} &\equiv \Avg{a_\mu,b_\nu} = \frac{1}{\nu!}
\int dx a_\mu(x)(-\partial)^\nu |\psi(x)|^2
\label{B}
\end{align}
is an invertible matrix. To prove so, apply integration by parts to
Eq.~(\ref{B}) to obtain
\begin{align}
B_{\mu\nu} &= 
\frac{1}{\nu!}\int dx  |\psi(x)|^2 \partial^\nu a_\mu(x)
= \sum_\xi A_{\mu\xi} C_{\xi\nu},
\label{B2}
\\
B &= A C,
\end{align}
where $C$ is defined by Eq.~(\ref{C}). Since $A$ is invertible, it
suffices to prove that $C$ is also invertible.  Consider the term
$\partial^\nu x^\mu$ in $C_{\mu\nu}$. $\nu > \mu$ in a
degree-respecting order implies $|\nu| > |\mu|$, or $|\nu| = |\mu|$
and $\nu \neq \mu$.  In either case, there exists at least one
$\nu_j > \mu_j$ that makes $\partial^\nu x^\mu$ vanish, resulting in
\begin{align}
  C_{\mu\nu} = 0\ \textrm{if}\ \nu > \mu,
\end{align}
meaning that $C$ is lower-triangular. The eigenvalues of $C$ are then
the diagonal elements and given by
\begin{align}
C_{\mu\mu} &= \frac{1}{\mu!}\int dx   |\psi(x)|^2 \partial^\mu x^\mu 
= \int dx |\psi(x)|^2 = 1.
\end{align}
Hence $C$ is invertible. Since both $A$ and $C$ are lower-triangular
and invertible, $B = AC$ is also lower-triangular and invertible, and
\begin{align}
B^{-1} = C^{-1}A^{-1}
\label{Binv}
\end{align}
is lower-triangular as well.

I can now use the $a$ basis to express Eq.~(\ref{Jdirectb}) as
\begin{align}
J_{\mu\nu} &= \frac{\tau}{\theta_0} \sum_\xi \Avg{b_\mu,a_\xi}\Avg{a_\xi,b_\nu}
= \frac{\tau}{\theta_0} \sum_\xi B_{\xi\nu} B_{\xi\nu}.
\end{align}
In matrix form,
\begin{align}
J &= \frac{\tau}{\theta_0} B^\top B,
\end{align}
and the CRB becomes
\begin{align}
\textrm{CRB} &= J^{-1} = \frac{\theta_0}{\tau} B^{-1} (B^{-1})^\top
\label{CRBdirect_B}
\\
&= \frac{\theta_0}{\tau} C^{-1} M (C^{-1})^\top,
\end{align}
where I have applied Eqs.~(\ref{Binv}) and (\ref{MA}).

\section{\label{sec_est}An unbiased and
efficient estimator for direct imaging}
Let $\{n(\mathcal S); \mathcal S \subseteq \mathbb R^D\}$ be the
Poisson process \cite{snyder_miller,cinlar} obtained by direct imaging
with infinitesimal pixel size. The expected value of $n$ over an area
$\mathcal S$ is
\begin{align}
\expect\Bk{n(\mathcal S)} &= \tau \int_{\mathcal S} dx f(x|\theta),
\label{mean_direct}
\end{align}
and $\{n(\mathcal S_1), n(\mathcal S_2), \dots\}$ are independent
Poisson variables if $\{\mathcal S_1,\mathcal S_2,\dots\}$ are
disjoint subsets. Consider the estimator
\begin{align}
\check\theta_\mu &= \frac{1}{\tau}\sum_\nu (C^{-1})_{\mu\nu} \int n(dx) x^\nu.
\label{est_direct}
\end{align}
Its expected value is
\begin{align}
\expect\bk{\check\theta_\mu} &= \sum_\nu (C^{-1})_{\mu\nu} \int 
dx f(x|\theta) x^\nu
\\
&= \sum_\nu (C^{-1})_{\mu\nu} \int dx \sum_\xi \frac{\theta_\xi}{\xi!}
(-\partial)^\xi |\psi(x)|^2 x^\nu
\\
&= 
\sum_\nu (C^{-1})_{\mu\nu} \sum_\xi \frac{\theta_\xi}{\xi!}\int dx 
|\psi(x)|^2 \partial^\xi x^\nu
\\
&= \sum_{\nu,\xi} (C^{-1})_{\mu\nu} C_{\nu\xi} \theta_\xi = \theta_\mu,
\end{align}
where I have applied Eqs.~(\ref{taylor}) and (\ref{C}).  Its
covariance, on the other hand, is
\begin{align}
\var_{\mu\nu}\bk{\check\theta} &= 
\frac{1}{\tau}
\sum_{\xi,\eta} (C^{-1})_{\mu\xi} (C^{-1})_{\nu\eta}\int  dx f(x|\theta) x^{\xi+\eta}
\\
&= \frac{\theta_0}{\tau} C^{-1} M (C^{-1})^\top,
\end{align}
which coincides with the CRB given by Eq.~(\ref{CRBdirect_exact}).
The estimator is hence unbiased and efficient.

\section{\label{sec_nodiff}CRB for direct imaging in
the  diffraction-unlimited regime}
Suppose that the PSF $|\psi(x)|^2 = \delta(x)$ is infinitely sharp
and $f(x|\theta) = F(x|\theta)$.  The image moments given by
Eq.~(\ref{M}) become identical to those of the object, viz.,
\begin{align}
M_{\mu\nu} &= \frac{\theta_{\mu+\nu}}{\theta_0},
\end{align}
the $C$ matrix given by Eq.~(\ref{C}) becomes
\begin{align}
C_{\mu\nu} &= \frac{1}{\nu!} \int dx \delta(x) 
\partial^\nu x^\mu = \delta_{\mu\nu},
\end{align}
and the CRB given by Eq.~(\ref{CRBdirect_exact}) becomes
\begin{align}
\textrm{CRB}_{\mu\nu} &= \frac{\theta_{\mu+\nu}}{\tau}.
\label{CRB_nodiff}
\end{align}
This represents an ideal scenario where the imaging is limited only by
shot noise and not by diffraction. Equation~(\ref{CRB_nodiff}) also
serves as a general lower bound on the CRB given by
Eq.~(\ref{fisher0}) for any linear-optical processing, as
Eq.~(\ref{p2}) is a Markov chain on $F(X|\theta)$ and the
data-processing inequality \cite{zamir} can be invoked.

To verify Eq.~(\ref{CRB_nodiff}), suppose that $F$ consists of
isolated point sources, viz.,
\begin{align}
F(X|\theta) = \sum_\sigma \vartheta_\sigma \delta(X-X_\sigma),
\end{align}
and since $|\psi(x)|^2 = \delta(x)$, their positions can be
perfectly resolved.  The unknowns are then $\vartheta$, and the CRB
with respect to $\vartheta$ is
\begin{align}
J_{\sigma\gamma}^{(\vartheta)} &= \frac{\tau}{\vartheta_\sigma}\delta_{\sigma\gamma},
&
\textrm{CRB}_{\sigma\gamma}^{(\vartheta)} &=
 \frac{\vartheta_\sigma}{\tau}\delta_{\sigma\gamma}.
\end{align}
Expressing the moments as
\begin{align}
\theta_\mu &= \sum_\sigma \vartheta_\sigma X_\sigma^\mu,
\end{align}
I can compute the CRB with respect to the moments 
via the transformation
\begin{align}
\textrm{CRB}_{\mu\nu} &= \sum_{\sigma,\gamma}\parti{\theta_\mu}{\vartheta_\sigma}
\textrm{CRB}_{\sigma\gamma}^{(\vartheta)}\parti{\theta_\nu}{\vartheta_\gamma}
= \frac{\theta_{\mu+\nu}}{\tau},
\end{align}
which coincides with Eq.~(\ref{CRB_nodiff}).

\section{\label{sec_matrices2} Properties of matrices in Sec.~\ref{sec_spade}}
Equation~(\ref{cpoly}) can be inverted to give
\begin{align}
k^r &= \sum_{s} (G^{-1})_{rs} g_s(k).
\end{align}
Substituting this in Eq.~(\ref{H}) and using the orthonormality given
by Eq.~(\ref{cortho}), I obtain
\begin{align}
H_{qr} &= \frac{i^{|q|}(-i)^{|r|}}{r!}\sum_{s} (G^{-1})_{rs} 
\int dk |\Psi(k)|^2 g_q(k) g_s(k)
\\
&= \frac{i^{|q|}(-i)^{|r|}}{r!}(G^{-1})_{rq}.
\end{align}
The inverse is given by Eq.~(\ref{Hinv}), which can be confirmed by
directly computing $HH^{-1}$ or $H^{-1}H$.  Since $G^{-1}$ and $G$ are
lower-triangular, $H$ and $H^{-1}$ are upper-triangular.

If $|\Psi(k)|^2$ is centrosymmetric according to Eq.~(\ref{even}),
Ref.~\cite{dunkl} shows that $g_q(k)$ consists of only even-order
monomials $\{k^r; |r| \ \textrm{even}\}$ if $|q|$ is even and only
odd-order monomials $\{k^r; |r| \ \textrm{odd}\}$ if $|q|$ is
odd. Thus
\begin{align}
G_{qr} &= 0 \ \textrm{if}\ |q|-|r|\ \textrm{is odd},
\\
g_q(k) &= (-1)^{|q|} g_q(-k).
\end{align}
Substituting $k$ with $-k$ in the integral in Eq.~(\ref{h_k}) yields
\begin{align}
h_q(X)&= i^{|q|}\int dk |\Psi(-k)|^2 g_q(-k)\exp(ik\cdot X)
\\
&= (-i)^{|q|}\int dk |\Psi(k)|^2 g_q(k)\exp(ik\cdot X)
\\
&= h_q^*(X),
\end{align}
and $h_q(X)$ is real. It follows that $H$ and $H^{-1}$ are real
as well.

\section{\label{sec_CRBgauss}CRB for direct imaging with the
Gaussian PSF}
In the limit of $\Delta \to 0$,
\begin{align}
\tilde f(x|\theta) &= |\psi(x)|^2 = \frac{1}{(2\pi)^{d/2}}
\exp\bk{-\frac{||x||^2}{2}}.
\end{align}
A set of orthogonal polynomials are
\begin{align}
a_\mu(x) &= \frac{1}{\sqrt{\mu!}}\He_\mu(x),
\label{He}
\end{align}
where
\begin{align}
\He_\mu(x) &\equiv \prod_{j=1}^D \He_{\mu_j}(x_j),
\end{align}
and the definition of the single-variable Hermite polynomials can be
found, for example, in Refs.~\cite{NIST:DLMF,Olver:2010:NHMF}.  The
$B$ matrix defined by Eq.~(\ref{B}) can then be computed by
substituting the identity
\begin{align}
(-\partial)^\nu |\psi(x)|^2 &= |\psi(x)|^2\He_\nu(x)
\end{align}
for Hermite polynomials \cite{NIST:DLMF,Olver:2010:NHMF} and using the
orthonormality of $a$. The result is
\begin{align}
B_{\mu\nu} &= \frac{1}{\sqrt{\mu!}}\delta_{\mu\nu},
\end{align}
which can be substituted into Eq.~(\ref{CRBdirect_B}) to give
Eq.~(\ref{CRBgauss}).

\section{\label{sec_nonsep}An estimator for SPADE with non-separable
  PSFs}
The simple estimator given by Eq.~(\ref{sim_est}) relies on the strong
upper-triangular property of $H$ given by Eq.~(\ref{strong_H}) for
separable PSFs.  Without it, the weaker property given by
Eq.~(\ref{weak_H}) for a degree-respecting order still implies that
the $\sum_r$ sum in Eq.~(\ref{theta_Gamma}) can be separated into a
$|r| = |q|$ group and and a $|r| > |q|$ group, viz.,
\begin{align}
\sum_r  &= \sum_{|r| = |q|} + \sum_{|r| > |q|},
\end{align}
and Eq.~(\ref{theta_Gamma}) becomes
\begin{align}
\theta_{q+q'} &= \sum_{|r|=|q|,|r'|=|q'|}(H^{-1})_{qr}(H^{-1})_{q'r'}\Gamma_{rr'}
\nonumber\\&\quad
+
\sum_{|r+r'|>|q+q'|}(H^{-1})_{qr}(H^{-1})_{q'r'}\Gamma_{rr'}.
\end{align}
If I assume the estimator
\begin{align}
\check\theta_{q+q'}' &= 
\sum_{|r|=|q|,|r'|=|q'|}(H^{-1})_{qr}(H^{-1})_{q'r'}\check\Gamma_{rr'},
\label{sim_est_nonsep}
\end{align}
the bias is also given by Eq.~(\ref{bias_mag}), while the variance is
\begin{align}
\var\bk{\check\theta_{q+q'}'} &= 
\sum_{|r|=|q|,|r'|=|q'|}\Bk{(H^{-1})_{qr}}^2\Bk{(H^{-1})_{q'r'}}^2
\nonumber\\&\quad\times
\var\bk{\check\Gamma_{rr'}}
\\
&= \frac{\theta_0^2}{\min(N_s)}O(\Delta^{|q+q'|}),
\end{align}
which can still be minimized by choosing $q$ and $q'$ according to
Eq.~(\ref{afloor}).

A problem with Eq.~(\ref{sim_est_nonsep}) is that, for a given $|q|$
and $|q'|$, the number of $(r,r')$ indices with $|r| = |q|$ and
$|r'| = |q'|$ is
\begin{align}
\bk{\begin{array}{cc}|q|+D - 1\\ |q|\end{array}}
\times
\bk{\begin{array}{cc}|q'|+D - 1\\ |q'|\end{array}},
\end{align}
so the estimator may require a large number of $\check\Gamma_{rr'}$'s
and a large number of bases to implement for a high-order moment,
leading to a reduction in $\min(N_s)$.  This difficulty is compounded
by the fact that, for $D \ge 2$, there exist infinitely many sets of
orthogonal polynomials for a given weight function, as pointed out in
Appendix~\ref{sec_CRBdirect}, leading to infinite possible choices of
the $g$ polynomials and the PAD basis. For separable PSFs, the choice
of the separable PAD basis in Sec.~\ref{sec_moment} fortunately leads
to only one term in Eq.~(\ref{sim_est_nonsep}), but it remains an open
question whether Eq.~(\ref{sim_est_nonsep}) can be further simplified
via a more specific choice of the PAD basis for non-separable PSFs.

\section{\label{sec_finiteM}Conditions for finite image moments}
Given Eqs.~(\ref{theta_mag}) and (\ref{Mseries}), $M$ is finite if all
the PSF moments $\{\Lambda_\mu; \mu \in \mathbb N_0^D\}$ are finite.
Consider
\begin{align}
\Lambda_\mu &= \int dk \Psi^*(k)(i\partial_k)^\mu \Psi(k)
\end{align}
in terms of the Fourier transform given by Eq.~(\ref{Psi}).  A
sufficient condition for $\Lambda$ to be finite is that $\Psi(k)$ is
infinitely differentiable and has compact support; an example is the
bump function given by Eq.~(\ref{bump}).

If any $\Lambda_\mu$ is infinite, the $C$ matrix given by
Eq.~(\ref{C2}) and the CRB given by Eq.~(\ref{CRBdirect}) also have
infinite elements, and the direct-imaging theory in
Sec.~\ref{sec_limits} and Appendix~\ref{sec_CRBdirect} breaks down.
This happens for the rectangle aperture function given by
Eq.~(\ref{hard}).  A solution, not explored in this work, may be to
smooth $\Psi(k)$ by convolving it with a bump function with support
width $w$, such that the smoothed $\Psi(k)$ becomes infinitely
differentiable but remains compactly supported. When $w \ll 1$, the
result should offer a good approximation of that for the original
$\Psi(k)$.


\bibliography{research}

\end{document}